%% file: output.tex
\PassOptionsToPackage{table}{xcolor}  

\documentclass[sigconf]{acmart}

\usepackage[table]{xcolor}  

\definecolor{lightgreen}{rgb}{0.88, 1.0, 0.88}  
\usepackage{lipsum}  
\usepackage{listings}
\usepackage[utf8]{inputenc}

\AtBeginDocument{%
  }
\copyrightyear{2025}
\acmYear{2025}
\setcopyright{acmlicensed}\acmConference[CHI '25]{CHI Conference on Human Factors in Computing Systems}{April 26-May 1, 2025}{Yokohama, Japan}
\acmBooktitle{CHI Conference on Human Factors in Computing Systems (CHI '25), April 26-May 1, 2025, Yokohama, Japan}
\acmDOI{10.1145/3706598.3714188}
\acmISBN{979-8-4007-1394-1/25/04}

\usepackage{multirow}
\usepackage{mathptmx}
\usepackage{CJKutf8}    

\usepackage{newtxmath}

\usepackage{longtable}
\usepackage{graphicx}
\usepackage{caption}
\usepackage{subcaption}
\usepackage{enumitem}

\usepackage{array}          

\begin{document}

\input{meta/commands}

\title{\systemname: Towards Proactive AR Assistant with Belief-Desire-Intention User Modeling
}

\input{meta/authors}

\begin{abstract}

\input{meta/abstract}

\end{abstract}

\input{meta/ccs}
\input{meta/keywords}
\begin{teaserfigure}
  \includegraphics[width=\textwidth]{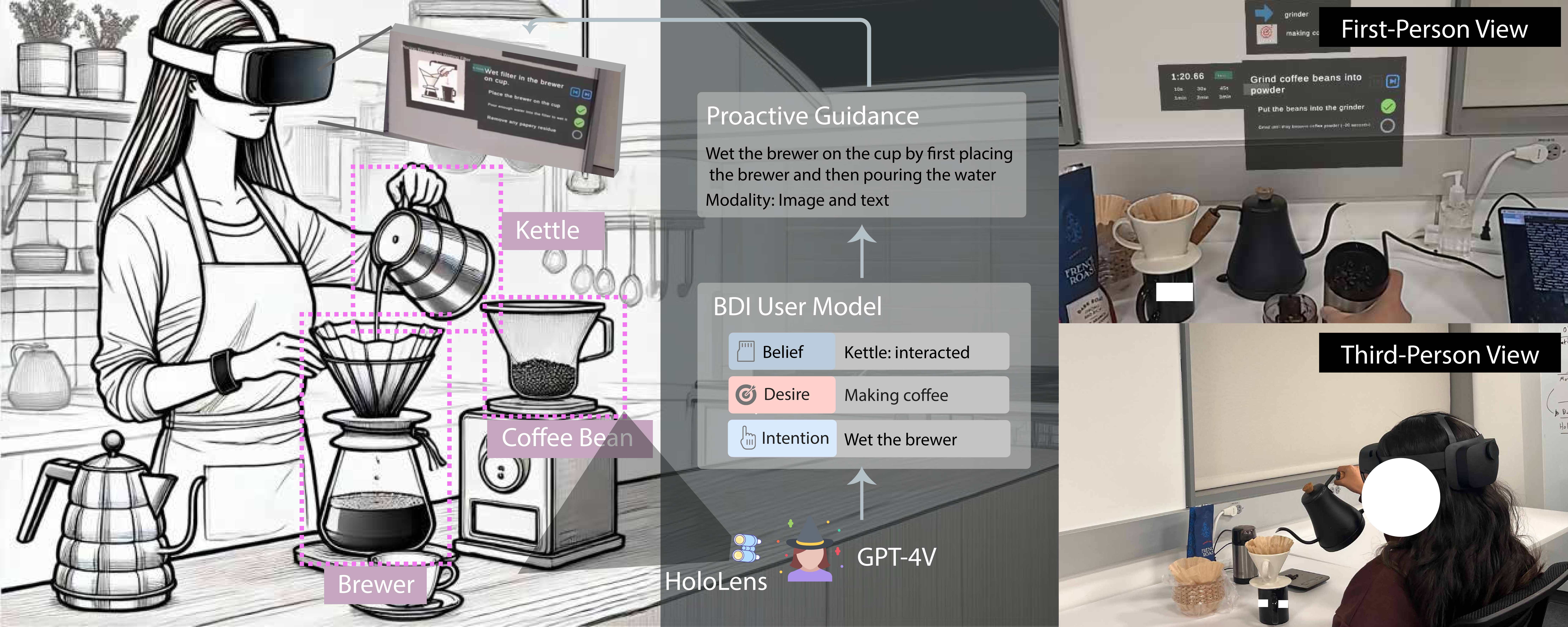}
  \caption{
    \systemname is a mind-reading monkey-shaped creature in Japanese folklore. Our system extends this concept to highlight the importance of incorporating the user's state (i.e., knowledge and intentions) while building proactive AR assistants. The Satori system combines the tracked objects, the surrounding environment, task goals, and user actions with a large-language model (LLM) model to provide AR assistance to the user's immediate needs. This kind of \textbf{proactive AR assistance} is achieved by implementing the Belief-Desire-and-Intention (BDI) psychological model with advice from two formative studies with a total of 12 experts. The \textit{belief} component reflects whether the users know where the task object is, and how to perform certain tasks (e.g., task goals, high-level knowledge); the \textit{desire} component is the \textbf{actionable goal}; and the \textit{intention} component is the \textbf{immediate next step} needed to complete the actionable goal. Our code is available at https://github.com/VIDA-NYU/satori-assistance.}
  \label{fig:teaser}
\end{teaserfigure}



\maketitle



\input{sections/01_introduction}

\input{sections/02_related}

\input{sections/03_formative}

\input{sections/04_method}
\input{sections/05_user}

\input{sections/05x_result}

\input{sections/06_discussion}

\input{sections/07_conclusion}
\bibliographystyle{ACM-Reference-Format}
\bibliography{output}

\appendix
\renewcommand\thefigure{\thesection.\arabic{figure}}



\end{document}

%% file: meta/commands.tex
\newcommand{\systemname}{Satori}
\newcommand{\baseline}{WoZ}
\newcommand{\gpt}[1]{\textcolor[HTML]{DDA0DD}{#1}} 

\newcommand{\mytodo}[1]{\colorbox{red!25}{\textcolor{white}{TODO: #1}}}

\newcommand{\todo}[1]{\textcolor{red}{[TODO] \emph{#1}}}

\newcommand{\textquote}[1]{\textit{#1}}
\newcommand{\concept}[1]{\textit{#1}}

\newcommand\myheading[1]{ \textbf{#1}}

\newcommand\revblue[1]{{#1}}

\newcommand\revisionbluetext[1]{\textcolor{blue}{#1}}
\newcommand\revision[1]{\textcolor{black}{#1}}

\newcommand{\chenyi}[1]{\textcolor[HTML]{00CFB6}{[Chenyi: #1]}} 
\newcommand{\guande}[1]{\textcolor[HTML]{FFC0CB}{[Guande: #1]}} 
\newcommand{\jing}[1]{\textcolor[HTML]{228B22}{[Jing: #1]}} 
\newcommand{\sonia}[1]{\textcolor[HTML]{DDA0DD}{Sonia: [#1]}} 
\newcommand{\gromit}[1]{\textcolor[HTML]{CC5500}{[Gromit: #1]}} 
\newcommand{\claudio}[1]{\textcolor[HTML]{008080}{[Claudio: #1]}} 

\newcommand{\participantquote}[1]{``\textit{#1}''}
\newcommand{\instructionquote}[1]{\textit{#1}}

%% file: meta/authors.tex
\author{Chenyi Li}
\authornote{Both authors contributed equally to this research.}
\email{chenyili@nyu.edu}
\affiliation{%
  \institution{New York University}
  \city{New York}
  \country{USA}
}

\author{Guande Wu}
\authornotemark[1]
\email{guandewu@nyu.edu}
\affiliation{%
  \institution{New York University}
  \city{New York}
  \country{USA}
}

\author{Gromit Yeuk-Yin Chan}
\email{ychan@adobe.com}
\affiliation{%
  \institution{Adobe Research}
  \city{San Jose}
  \state{California}
  \country{USA}}

\author{Dishita G Turakhia}
\email{d.turakhia@nyu.edu}
\affiliation{%
  \institution{New York University}
  \city{New York}
  \state{New York}
  \country{USA}
}

\author{Sonia Castelo}
 \email{s.castelo@nyu.edu}
\affiliation{%
 \institution{New York University}
 \city{New York}
 \country{USA}}

\author{Dong Li}
  \email{dl5214@nyu.edu}
\affiliation{%
  \institution{New York University}
  \city{New York}
  \country{USA}}

\author{Leslie Welch}
\email{leslie_welch@brown.edu}
\affiliation{%
  \institution{Brown University}
  \city{Providence}
  \state{Rhode Island}
  \country{USA}}

\author{Cl\'{a}udio T. Silva}
\authornote{Co-corresponding authors}
\email{csilva@nyu.edu}
\affiliation{%
  \institution{New York University}
  \city{New York}
  \country{USA}}

\author{Jing Qian}
\authornotemark[2]
\email{jq2267@nyu.edu}
\affiliation{%
  \institution{New York University}
  \city{New York}
  \country{USA}}

\renewcommand{\shortauthors}{Li and Wu, et al.}

%% file: meta/abstract.tex
Augmented Reality (AR) assistance is increasingly used for supporting users with physical tasks like assembly and cooking. However, most systems rely on reactive responses triggered by user input, overlooking rich contextual and user-specific information. To address this, we present Satori, a novel AR system that proactively guides users by modeling both -- their mental states and environmental contexts. Satori integrates the Belief-Desire-Intention (BDI) framework with the state-of-the-art multi-modal large language model (LLM) to deliver contextually appropriate guidance. Our system is designed based on two formative studies involving twelve experts. We evaluated the system with a sixteen within-subject study and found that Satori matches the performance of designer-created Wizard-of-Oz (WoZ) systems, without manual configurations or heuristics, thereby improving generalizability, reusability, and expanding the potential of AR assistance.  Code is available at https://github.com/VIDA-NYU/satori-assistance.


%% file: meta/ccs.tex
\ccsdesc[500]{Human-centered computing~Mixed / augmented reality}
\ccsdesc[500]{Human-centered computing~User models}

%% file: meta/keywords.tex
\keywords{Augmented reality assistant, proactive virtual assistant, user modeling}

%% file: sections/01_introduction.tex
\section{Introduction}

\systemname~(
\begin{CJK}{UTF8}{gbsn}
悟
\end{CJK}
\begin{CJK}{UTF8}{min}
り
\end{CJK})
, a ghost-like deity from Japan, is fabled to read human minds and respond to thoughts before they unfold into action. While such supernatural power once belonged strictly to the realm of folklore, modern AI technologies are now beginning to emulate a similar ability of predicting human intent and actions and even providing proactive assistance during task interactions~\cite{DBLP:conf/icmi/KrausSBBDDGBM20}. Such \textit{proactive} virtual or digital assistance, which determines optimal \concept{content} and \concept{timing} without explicit user commands, is gaining traction for its ability to enhance productivity and streamline workflow by anticipating user needs from context and past interactions~\cite{DBLP:journals/corr/abs-2005-01322}. However, there is currently limited research on how to best design and implement such systems. 

Most current assistance in augmented reality (AR) remains \textit{reactive}, responding to user commands or environmental triggers without the capacity for \textit{active} engagement. These systems require that users initiate interactions, which is inefficient in AR where users typically have limited attention to spare. In response to this, some AR assistance incorporate proactive elements; for instance, they may provide maintenance guidance based on recognized objects or components~\cite{lai2020smart, reflective-ar, DBLP:journals/imwut/MeurischMHGMHM20}. Yet, these systems are often built on fixed rules and lack adaptability and reusability. \revision{They are limited in responding effectively to the user's surrounding environment or interpreting their actions over time. As a result, these systems struggle to guide users across multiple, consecutive steps and instead tend to function as discrete task-only assistance.}

Designing proactive assistance for AR is particularly challenging due to the necessity of understanding the user's state, short-term goals, and surrounding environment. Further, timely assistance is crucial due to constraints on user attention. Providing assistance too early, too late, or simply too frequently can increase cognitive load and negatively impact the user's experience~\cite{baumeister2017cognitive, adaptive21}.  

In this paper, we address these gaps by first identifying the in-depth benefits, challenges for designing a proactive AR assistance by conducting two formative studies, and then exploring the design of a system through \systemname. The first study with six professional AR designers revealed several design challenges such as: 1) limited generalizability and reusability of current non-proactive AR assistance, 2) difficulties in accurately detecting user intentions, and 3) the need to balance general advice with task-specific solutions. The professionals recognized that using proactive AR assistance could potentially improve
scalability and efficiency, but also highlighted the technical challenges related to accurately tracking and understanding users' actions.

Building on the findings from the first study, the second formative study engaged six experts---three human-computer interaction (HCI) researchers and three psychology researchers---in dyadic interviews to explore design strategies for a more proactive AR assistance. The design sessions found four key design considerations: 1)  understanding human actions; 2) recognizing surrounding objects and tools; 3) assessing the current task; and 4) anticipating immediate next steps. Following experts' suggestions, these findings were later integrated with the well-established belief-desire-intention (BDI) model~\cite{georgeff1999belief, leslie2005belief, bratman1987intention, cohen1990intention}, resulting in an AR-specific adaptation that guided the development of our system, \systemname.

To adapt the BDI model for AR assistance, \systemname~needed to \revision{address the challenges brought up in the formative studies and} account for the limitations of the AR headset. Inspired by the theory underpinning the BDI model, we build \systemname~using an ensemble of egocentric vision models combined with a multimodal large language model (LLM) to determine timing, content, and user action in everyday AR assistance. The system is a multi-modal proactive assistance wherein the user's environment, nearby physical objects, action history, and task goals are input to predictively determine the content and timing of the assistance. Our approach ensures that the AR assistance delivers relevant information at appropriate moments, enabling a new and more seamless experience for AR users.

We evaluated \systemname~\revision{over four everyday AR tasks and compared it to a Wizard-of-Oz system (i.e., baseline) designed by six professional AR designers. We found that \systemname's proactive guidance was as effective, useful, and comprehensible as the AR assistance created by the designers. User ratings also indicated that \systemname's timing prediction performs similarly to the baseline}. Additionally, \systemname's guidance allowed participants to switch between tasks without the need for pre-training or scanning. Our findings suggest that our application of the BDI model not only successfully understood users' goals and actions but also captured the semantic context of given tasks, reducing the need to craft AR assistance for every specific scenario and improving its generalizability and reusability. 

To summarize, our contributions include:
\begin{enumerate} 
\item  Identifying benefits, challenges, and design requirements for creating a proactive AR assistance, derived from two formative studies with twelve experts and applied using concepts from the BDI model in AR environments.
\item Design and implementation of \systemname, a proactive AR assistance system applied with BDI's concepts that combine LLM with a series of vision models to infer users' current tasks and actions, providing appropriately timed step-by-step assistance with dynamically updated content. 
\item A 16-user empirical study shows that \systemname~delivers performance comparable to designer-created AR assistance in terms of timing, comprehensibility, usefulness, and efficacy.
\end{enumerate}

%% file: sections/02_related.tex
\section{Related Work}

Virtual assistants in Augmented and Virtual Reality (AR/VR)  can well support tasks in assembly and manufacturing~\cite{DBLP:conf/vr/BottoCCMSRDS20, konig2019ma, reflective-ar, LieVR}, surgery~\cite{qian2018arssist, escobar2020review}, maintenance~\cite{konstantinidis2020marma, frandsen2023augmented, borro2021warm} and cooking~\cite{DBLP:conf/metroi/DAgostiniBSPFLM18}. Such systems are often task-specific and their principles are not easily generalizable across varied domains. One way to improve generalizability is via a command-based AR assistant, which can enhance user confidence in the system's real-world awareness ~\cite{kim2018does}. Yet, command-based assistance requires the user's explicit input and thus limits usability. Our work builds on previous research related to virtual assistants in AR/VR , while addressing user needs without explicit commands or domain limitations.

Predesigned assistance in AR and VR applications typically involves preset rules for actions or reminders triggered by specific user inputs or situations. This kind of rule-based assistance is simple and intuitive, providing users with readily available support that can be accessed on demand or in time sequences~\cite{10.1007/978-3-031-06015-1_21}. While implementing such assistant systems is straightforward~\cite{sara2022assessment}, they require extensive manual user interaction to describe and confirm the user's needs. For example, Sara et al. demonstrated an AR maintenance assistant wherein the technician needed to manually confirm the completion of each step and proceed to the next step using touchpad controls or voice commands~\cite{sara2022assessment}.  

Proactive assistance, on the other hand, is designed to actively recognize context information and infer user intentions even if they are not explicitly provided ~\cite{10.1145/3624918.3629548, 4415181, 9680771}. Such assistance is designed without human intervention~\cite{DBLP:conf/icmi/KrausWM21, DBLP:journals/spm/Sarikaya17, DBLP:conf/cui/ZarghamRBVSMR22}, aiming for tasks across different domains, such as health care~\cite{DBLP:conf/huc/RabbiAZC15, DBLP:conf/huc/SchmidtBEM15}, navigation~\cite{pejovic2015anticipatory} and laboratory education~\cite{DBLP:conf/huc/SchollWL15}.
It enhances usability~\cite{DBLP:conf/lrec/SchmidtMW20}, fosters trust~\cite{DBLP:journals/access/KrausWCM21}, and improves task efficiency~\cite{DBLP:journals/ijait/Yorke-SmithSMM12}. During AR interactions, proactive assistance often takes the user's surrounding environment into account, predicts the user's goals, and offers context-aware recommendations, often for the sake of improving attention~\cite{DBLP:conf/huc/MeurischISM17, jain2023co, shih2020digital, meck2023may}. \revision{For example, gaze-moderated systems such as iBall demonstrate how gaze data can be integrated into visualizations to enhance task engagement and attentiveness~\cite{chen2023iball}} However, existing proactive assistance relies on preset rules such as location, time, and events to trigger the assistant's intervention~\cite{DBLP:conf/huc/MeurischISM17}. For instance, Ren et al. \cite{REN2023963} propose a proactive interaction design method for smart product-service systems, using functional sensors to gather explicit contextual data (e.g., physical location, light intensity, environment temperature) to predict implicit user states (e.g., user attention level). Although these methods advance the progress of proactive assistance, such signals may not align with the actual users' needs, leading to ineffective and obtrusive assistance~\cite{DBLP:conf/interact/XiaoCS03, DBLP:conf/lrec/KrausWM22}. To address this, we propose using the user's intention, goals, and the interaction context to dynamically determine the assistance's timing, content, and modality.

Most AR assistance today remain passive because defining user intention is difficult. One challenge is that understanding users' intention relies not only on explicit cues (e.g., verbal statements or signals) but also, significantly, on implicit non-verbal cues and body postures~\cite{kim2018does}. Successfully decomposing and reasoning with implicit cues improves the chances of intention labeling.  Recent advancements in vision-language models offer new opportunities to integrate body postures into AR assistance. Therefore, we propose a multimodal input mechanism that uses voice and visual cues to better understand users' intentions.

In the following subsection, we discuss the latest related work on the various aspects of designing proactive task guidance in AR, such as predicting egocentric actions, understanding user intention, building models on the theory of mind and BDI framework, and modeling human-AI collaboration. 

\revision{
\subsection{Egocentric Action Prediction}
Egocentric action prediction focuses on forecasting users' future actions or interactions based on first-person video data, leveraging temporal and multimodal cues. Recent works have explored multimodal approaches, such as transformer-based architectures, to integrate visual and contextual information for early action prediction~\cite{guan2023egocentric}. Similarly, intention-based models have been proposed to emulate human-like reasoning in predicting future object interactions in egocentric settings~\cite{ma2024egocentric}. Surveys highlight the growing importance of egocentric data for understanding human actions and intentions due to its unique ability to capture users' perspectives directly~\cite{kong2022human, rodin2021predicting}. Our system extends these approaches by incorporating a dual-modal analysis of visual and semantic cues to provide adaptive and real-time action anticipation.
}









\subsection{Understanding User Intention}
Understanding user intention is paramount for improving users' interaction and experience with electronic devices, ranging from smart mobile devices to augmented reality (AR) systems. Research in the field of information needs has highlighted the importance of intention classification and systematic taxonomy in achieving this goal. Border proposed a taxonomy of web searches, classifying intentions into navigational, informational, and transactional~\cite{broder2002taxonomy}. This groundbreaking work laid the foundation for more detailed classifications. For instance, Dearman et al. categorized sharing needs and sharing entries into nine distinct categories, extending the concept of information needs to a collaborative level~\cite{dearman2008examination}. This classification allows developers to design products that better facilitate collaborative information sharing. Church et al. found that contexts such as location, time, social interactions, and user goals influence users' information needs. For example, it was found that users generated more locational or temporal dependencies when they were on the go. Users also require more geographical information when they are commuting. This study enabled researchers to design an information search platform, SocialSearchBrowser, to fit users' unique information needs in a context-sensitive way~\cite{church2009understanding}. Additionally, Li et al. extended this research by developing a design space of digital follow-up actions~\cite{li2024omniactions}. They classified actions into 17 types and identified seven categories of follow-up actions through a qualitative analysis of users' diaries. They also deployed the system on mobile AR and conducted a user study to test the capacity of follow-up action designs. The study showed that the system could accurately predict users' general actions, provide proactive assistance, and reduce friction~\cite{li2024omniactions}. Generally, prior studies on information needs, particularly on mobile devices, have demonstrated that intention taxonomy could inspire the design of information search systems with more proactive and contextual assistance.

\subsection{Theory of Mind}
\revision{
Theory of Mind (ToM) refers to the ability to understand \revision{user's} mental states, such as beliefs, intentions, and desires, and to use this understanding to predict and interpret behavior. The concept, first introduced by Premack and Woodruff~\cite{premack1978does}, has become a cornerstone of cognitive psychology and neuroscience. ToM is recognized as critical for social interaction and communication, enabling individuals to navigate complex social environments~\cite{frith2005theory, carlson2013theory}. 
}
\revision{The development of ToM progresses from implicit understanding in infancy to explicit reasoning about mental states during early childhood~\cite{wellman2018theory, blijd2017non}. This developmental trajectory highlights its reliance on both domain-general cognitive processes, such as executive function, and domain-specific skills, like language~\cite{bamicha2022evolutionary}. Recent advancements in artificial intelligence have sought to emulate ToM for applications in social AI and potentially in AR/VR systems. For instance, Wu et al. introduced COKE, a cognitive knowledge graph that formalizes ToM through structured cognitive chains, illustrating its potential for enhancing machine understanding of human social behavior~\cite{wu2023coke}.
}

\subsection{Belief-Desire-Intention Framework}
The Belief-Desire-Intention (BDI) model \cite{leslie2005belief, kahneman2013prospect, bratman1987intention, cohen1990intention} is a framework to simulate human decision-making behaviors in both individual~\cite{rao1997modeling} and multi-agent settings~\cite{pauchet2007computational, kim2012modeling, lee2009integrated}. The model originates from folk psychology and is extensively applied in cognitive modeling, agent-oriented programming, and software development. This model comprises three primary components: beliefs, desires, and intentions ~\cite{bratman1987intention}. Beliefs represent the information that humans perceive about a situation (e.g., It is raining), limited by their perceptions. Desires are the goals that individuals aim to achieve given the current situation (e.g., A person prefers not to get wet during a rainy day). Intentions are ``conduct-controlling pro-attitudes, ones which we are disposed to retain without reconsideration, and which play a significant role as inputs to [means-end] reasoning''~\cite{bratman1987intention}. In other words, the user's behavior moves toward achieving the desire (i.e., goal) by selecting and committing to specific plans of action (e.g., planning to get an umbrella).

Previous studies have demonstrated the effectiveness of the BDI framework in modeling human behavior~\cite{pauchet2007computational, kim2012modeling}. Therefore, the BDI model can help in the building of intelligent agents in various applications. For example, in agent-oriented programming, the BDI model is pervasively used to model an agent executing programming functions. Agent-oriented software engineering utilizes beliefs, actions, plans, and intentions to develop programs. The BDI model enables more rational and autonomous executions in unpredictable environments, such as AgentSpeak(L)~\cite{rao1996agentspeak}, 3APL~\cite{hindriks1998formal}, JACK~\cite{busetta1999jack}, JADEX~\cite{braubach2005jadex}, and GOAL~\cite{hindriks2009programming}. One benefit of using the BDI framework is that it makes agent behavior intelligible to end users and stakeholders. By committing to specific courses of action or intentions, BDI agents enhance user understanding and the predictability of actions~\cite{bordini2020agent,rao1998decision, kinny1996methodology, karaduman2023rational, fichera2011flexible, araiza2016model, gottifredi2008bdi, duffy2005social, pereira2005towards, ujjwalcase}.

Though BDI-inspired agents have enabled automatic decisions, making decisions in AR requires a different type of intelligent and realistic behavior. The environment for AR applications involves complex real-world dynamics, such as egocentric video, audio, and gestural inputs~\cite{bohus2024sigma}. The users' interaction goals, physical actions, and surrounding context (e.g., objects, tools, interaction agents) further increase the difficulty of providing in-time assistance~\cite{lindlbauer2019context}. Although the BDI framework has not yet been applied to AR, our work draws inspiration from the philosophy and design of prior BDI-based systems to enhance AR assistance. With recent advancements in LLMs, BDI-driven agents present a promising direction~\cite{bordini2020agent}, as LLMs can naturally serve as interpreters and reasoning machines, bridging language and text within the BDI framework.

\subsection{User Modeling in Human-AI Collaboration}
Modelling the user state is a long-standing problem in HCI \cite{DBLP:journals/air/McTear93a, DBLP:journals/kbs/BenyonM93}. Previous research focuses on the user goal and intent \cite{DBLP:journals/jiis/Yadav10}, expertise modeling to support adaptive computing systems \cite{DBLP:journals/hhci/VaubelG90}, and the study of the memory of the user for AR/MR-specific research \cite{DBLP:journals/corr/abs-2308-05822, DBLP:journals/percom/HarveyLW16}. The BDI model, a commonly accepted psychological framework \cite{leslie2005belief, georgeff1999belief}, becomes crucial in the emergent human-AI collaboration, necessitating a better model of the user state \cite{DBLP:conf/hicss/LaiKO21}. Existing research, however, focuses on the user's intention and goal and seldom addresses the user's knowledge or belief \cite{DBLP:conf/cvpr/WuLS22, DBLP:conf/chi/0002LY22, wu2024your, DBLP:journals/corr/abs-2308-16785, Turakhia2024Generating}. Furthermore, there's a lack of distinction between high-level goals (desires) and immediate goals (intents) \cite{koo2016structural}. Hence, we propose a general model for the user state, amalgamating belief, desire, and intent.

%% file: sections/03_formative.tex
\section{Formative Study 1: Design with Professional AR Designers}
We first conduct a formative study to explore the problem space and potential benefits of proactive AR assistance. The study begins with a semi-structured interview on participants' background knowledge, followed by designing four different common AR interaction scenarios. A final apparatus combining participants' design feedback is created for later study. 

\subsection{Participants}
Using email and snowball sampling, we recruited six professional AR designers (three female and three male, age: $\bar{x} = 30$). As we wanted to collect insights from experienced individuals, all participants selected were professional with at least three years of experience of working on developing AR applications. Participants were paid $\$30$ per hour.

\subsection{Tasks}
The study was conducted in two sessions: a semi-structured interview and a design session for four different everyday AR scenarios with assistance. Each participant was asked to design two out of the four scenarios for a balanced scenario distribution. Each scenario was designed by three different AR designers. 

In the first session, we collected participants' prior working experience using AR assistants, the challenges they faced in creating them, and their assessment of the assistants' potential benefits and applications. Additionally, we discussed the concept of proactive AR assistance with participants and collected their insights on potential benefits and use scenarios.
In the second session, participants were asked to design AR assistants for two everyday scenarios out of the four. These two scenarios were assigned in a pre-determined order to balance the total number of designs. We use WikiHow~\footnote{https://www.wikihow.com/} to obtain detailed, step-by-step instructions as the \textbf{task background information} for participants. These instructions ($average steps: \bar{x}=7$) provide the framework to make guidance, and participants can elaborate (e.g., adding additional steps) at their will. Aside from the text instructions, we recorded videos in first-person view using the original instructions to provide visual reference and interaction context for participants. 
Given instructions, images, and videos depicting the scenarios, participants were asked to design: 1) if a piece of guidance is needed for a particular step; 2) when the guidance should appear and for how long; 3) the modality of the guidance; 4) the content of the guidance. The above questions focus on the questions of ``if'', ``when'', ``how'', and ``what'' in AR assistance, which is a common architecture for guiding users in the literature and current practice~\cite{lindlbauer2019context}.

\subsection{Procedure}
Since the AR designers reside in different time zones, the experiment was conducted remotely via Zoom after obtaining their informed consent. Participants were asked to introduce their background, describe their daily work, and discuss their projects related to AR assistance. We further inquired about their insights into the advantages and disadvantages of AR assistance, including challenges faced during development and challenges faced by end users. Finally, we presented the concept of proactive AR assistance and solicited their opinions on potential challenges and applications, as well as feasibility. 

After the semi-structured interviews, participants received digital forms containing materials to design AR assistance for their assigned tasks, including textual descriptions and contextual images and videos. During this phase, participants were introduced to the interface and how to use its operations to, for example, create interaction prompts for a step/sub-step or select what information the user should be presented with in what modality.  The experimenters addressed any questions participants raised via Zoom.

On average, the study's first session lasted approximately 28 minutes ($\bar{x}$ = 28), while the second session took around 60 minutes ($\bar{x}$ = 60), totaling around 90 minutes. All participants successfully completed the design task. Since every scenario was designed twice by two participants, the final AR assistance design was merged together by the experimenters based on common modalities and union of participant-generated instructions. Inconsistencies were resolved through discussion.

\subsection{Results}
\subsubsection{Benefits in conventional AR assistance}
\paragraph{AR assistance is beneficial in providing real-time, contextual information that improves user awareness}. Such guidance has ability to reveal forgotten or overlooked information. For instance, P1 emphasized that \participantquote{I find AR assistance most useful when it helps the user realize something they might not know... they might forget about an object, or not be aware that this object could be used in this situation... then (with AR assistance) they have this eureka moment. } 

\paragraph{AR assistance is also typically intuitive for users to follow, which reduces interaction cost and supports decision-making. } P2 and P3 highlighted that by overlaying visual cues such as arrows or animations directly onto the environment, AR could help users quickly comprehend otherwise difficult tasks such as examining electrical circuits. P3 stated that \participantquote{in tasks with spatially sensitive movements... AR is a proper medium because users intuitively understand what they need to do.} P3 further explained that users who received spatially directed AR guidance for operating a machine (e.g., turning knobs or pressing buttons) found it more intuitive than 2D instruction books or manuals. Additionally, P4 brought up that being able to provide spatial guidance reduces interaction costs for tasks that require frequent operations, simplifying users' decision-making process.


\subsubsection{Challenges in conventional AR assistance}
\paragraph{Pre-designed AR assistance are hard to scale to diverse contexts.}
AR designers often create designs based on their assumptions about the user's environment. However, users may interact with objects that fall outside these initial assumptions. As P1 noted, \participantquote{It's hard to cover all the edge cases of what a person might have... I assume they're in an indoor space, but that might not be the case,} highlighting the complexity of accommodating varied environments. 

\paragraph{AR assistance lacked an interaction standard.} P5 noted that there is not a standardized approach in the expansive interaction design space, especially when compared to traditional 2D interaction.  P3 expressed that creating 3D visual assets from scratch was usually complicated.

\paragraph{Predicting action timing and user intention remains challenging. } Both P3 and P4 noted the difficulty in defining an accurate mapping between user actions and AR responses. P4 emphasized that misinterpreting user behavior can result in irrelevant or unhelpful guidance (e.g., recommending a taxi when the user intends to walk). P3 also emphasized the difficulty faced by task experts who do not have engineering expertise, stating, \participantquote{Suppose I am a designer and I know nothing about coding, but I still want to make AR assistance for users. How should I do that?}

\subsubsection{Benefits of Proactive AR Assistance}

\paragraph{Proactive AR assistance is automatic without needing user input.} During the later-part of the interview, participants envisioned potential benefits of applying proactive AR assistance on common tasks, from both the AR developers' and users' point of view. Three participants described an automatic AR assistance as \textbf{proactive assistance} as P4 pointed out that such assistance anticipates the user's intentions and actively provide guidance based on the user's surrounding environment.
\paragraph{Reduces development time and increases efficiency.} Half of participants (P1, P2, and P6) agreed that proactive assistance could tremendously reduces development time on similar AR assets, animations, and programming logic (e.g., a panel shows up when a user touches an {object}). For instance, P2 remarked, \participantquote{We will definitely see a huge improvement in the efficiency of the content creation through this auto-generation process.} P1 said that automatic assist can simplify the repetitive design process in \participantquote{adding labels, recognizing objects, and generating guidance}. She continued to offer an example of a cooking app where such automation would be particularly useful in identifying ingredients or suggesting cooking steps. 

\paragraph{Improve scalability.} Both P1 and P3 highlighted how automatic AR assistance could generalize across different domains. According to P1, \participantquote{If we have a pipeline... using computer vision, it would save a lot of time... could have a universal pipeline to create guidance.} Moreover, P3 pointed out that such assistance may be adapt as authoring tools like spatial programming and program-by-demonstration, increasing the accessibility for non-developer users. 

\paragraph{Reducing information overload.} Participants (P3, P5) pointed that proactive assistance could automatic detects user's intention during AR interaction, presenting live-updated information in-need, thus reducing information overload. It may also gain trust from users since the proactive assistance might make users believe that the system understand their intentions well.


\subsubsection{Challenges of Proactive AR Assistance}
\paragraph{Cross-domain scalability is difficult.} P1 raised concern over the feasibility of a universal system that could operate across different devices and domains.  P3 further added that scalability remains a primary hurdle even for the most experienced AR designers because domain-specific knowledge is usually required to provide effective guidance. \participantquote{Scalability is the main issue... AR systems must lie in a specific domain, and it's hard to do this for every domain.} P6 brought up the fact that proactive assistance must be able to adapt to even unforeseen circumstances, which requires a deep understanding of the task at hand. Even with the help of LLMs, further training and customization of the tasks have been necessary, as LLMs are generally not domain-specific.

\paragraph{Detecting user intention is a primary challenge, as errors lead to confusion.} Four participants (P2, P4, P5, and P6) emphasized the difficulty of accurately detecting users' intentions in AR. P5 brought up the limited field of view (FoV) in AR headsets and the low accuracy of detection algorithms as two main issues, although the former (limited FoV) might be among the causes of the latter (low accuracy). P5 commented that \participantquote{...sometimes, the system might trigger guidance when the user doesn't need it, which could lead to confusion...} Similarly, P4 discussed how AR software in the industry has struggled to fully apprehend complex user environments and actions, causing confusion. This view is also shared by P2, who mentioned that proactive assistance might confuse users if it lacks self-explanatory features. P2 stated that \participantquote{if (the system is) fully automatic, you need the system to have some type of feedback. Automation without feedback may confuse the user.}

\paragraph{Adapting the AR instruction to users' active duties is challenging.} P6 stressed that a proactive system should automatically adjust general advice to task- (and) environment-specific solutions. AR systems must remain relevant to the user's current goal,  offering guidance that is actionable and appropriate.  


\subsubsection{Design results for four common scenarios}
All participants used user-centered and object-centered strategies to determine when assistance should appear. Participants using the user-centered strategy focused on actions by, for example, \textit{showing an instruction when the user got stuck on a step} or was about to get stuck. They also created instructions to indicate the user's completion of a step or unexpected situations. Participants who were conversely focused on object-centered strategies designed AR assistance that appeared in response to objects of interest. For example, one participant designed a reminder to \textit{change the mop pad} when \textit{the old pad is dirty}.
 
Participants' designs comprised multiple modalities, such as text, visuals, audio, and sometimes even tools (e.g., a timer). Notably, they tended to combine modality (``how'') with specific contents (``what''), see Table ~\ref{tab:modality_assistance}. While most participants chose to use text-based assistance to provide an overview of step-by-step instructions, information about the object, or reminders, they also designed three types of visuals: overlays (e.g., arrows, progress indicator, checkpoint cue), images, and animations. In addition, audio was repeatedly used to sound a warning, pronounce guidance, or indicate completion.

\begin{table*}[ht]
\begin{tabular}{lll}
\toprule
Modality    & Detailed Assistance Type  & Content                            \\ \midrule
text        & text                      & overview; instruction information; reminder \\
visuals     & animations                & instruction                                \\
            & image                     & instruction                                \\
            & arrows                    & location; interaction point                \\
            & checkpoint cue            & step completion; warning                   \\
audio       & sound cue                 & step completion; warning                   \\
            & voice                     & instruction                                \\
tools       & timer                     & count time                                 \\ \bottomrule
\end{tabular}
\caption{Types of assistance provided across different modalities suggested by expert AR designers. The overlays are used to indicate locations or to indicate how to interact with apparatus in the scene; a progress indicator reflects how far the user is into the task. The image and animation are designed to illustrate actions and positions and show ``how'' to complete the current step. The checkpoint cues, according to participants, are used to indicate step completion. The timer counts time for time-sensitive steps, such as making pour-over coffee.}
\label{tab:modality_assistance}
\end{table*}

\subsubsection{Wizard-of-Oz system}
Each participant created two AR assistance designs for two distinct tasks, totaling to 12 designs for four tasks. These designs were later combined into a Wizard-of-Oz (WoZ) system. The system contains in-situ image, voice, and text-based AR assistance displays. We combined similar timing, modality, and content to form one AR assistance per task. Images were sourced from task instructions on WikiHow, and text and voice guidance were developed by combining participant designs and WikiHow instructions. We then implemented the four AR assistance architectures in Unity and employed WoZ to trigger the assistance on time and accurately via a wireless keyboard controlled by a human experimenter. To visualize instructions, we overlaid them directly on static images to indicate where the interaction should happen, how many materials should be used, etc. Animations were achieved by looping multiple image sequences, similar to a GIF animation. The resulting system was video-recorded over Microsoft HoloLens and sent back to participants for recognition. All agreed with how each step was implemented after discrepancies were resolved either through clarification or modification of the apparatus.

\section{Formative Study 2: Co-Design with Psychological and HCI Experts}

\label{sec:codesign}
Building on the previous formative study, the second formative study sought to gain insights into the design of a proactive AR system by consulting experts. We recruited six experts, three from computer science and three from psychology (E1-6). \revision{The study focused on \textbf{how to design} the system and \textbf{the probable methods} for executing said design by discussing critical factors, interaction flows, and system architectures via two dyadic interviews.} We paired experts with complementary backgrounds to form three groups (Groups A, B, and C) as Table~\ref{tab:expert} shows. Their ideas and designs motivated later system implementations.

\subsection{Dyadic Interviews}
During the dyadic interviews, each pair of participants worked together to respond to open-ended questions and goals~\cite{morgan2013introducing}. \revision{ The first interview incorporated \textit{participatory design} to explore potential solutions; the second interview focused on designing detailed interaction flows and system architecture.} During the first interview, a set of goals and known challenges were presented to the groups to establish context; we included common AR assistance scenarios such as kitchen food preparation, classroom education tasks, and factory workflows. 

\subsection{Known Challenges}
\label{sec:codesign:background}
We presented participants with known challenges drawn from two sources, a literature survey and the results of the first formative study. The literature survey, which was furnished by searching \textit{AR assistance, embodied assistant, and immersive assistant} on Google Scholar and ACM DL, is described in the following subsections. Two authors separately reviewed these papers, coded the challenges, and formed themes from the coding. In total, 25 common challenges were identified and grouped using thematic analysis~\cite{braun2012thematic}.

\subsubsection{\textbf{C1}: Triggering assistance at right time is challenging.} 
AR assistance must be triggered at the appropriate time during AR interaction. Poor timing strategy may confuse users and negatively impact user trust~\cite{DBLP:conf/um/KrausWM20}. If a user is occupied or under stress, for example, frequently or inappropriately displaying AR assistance may be distracting or compound stress. Existing practice in AR assistance regulates the timing and display frequency using the user's intent and actions~\cite{DBLP:journals/spm/Sarikaya17} or fixed intervals. However, these methods do not consider the user's goal and lead to sub-optimal performance. 

\subsubsection{\textbf{C2}: Reusability and scalability in AR assistance are a problem.}
Most existing AR assistance systems are designed with ad hoc solutions, where the assistance (e.g., image, text, or voice) is individually developed~\cite{DBLP:conf/huc/RabbiAZC15, pejovic2015anticipatory, DBLP:conf/huc/MeurischISM17} and later adapted for re-use because each interaction scenario is likely to be unique. This creates repetitive labor, a concern raised by professional AR designers in our previous formative study.

\subsubsection{\textbf{C3}: Task interruption and multi-task tracking pose challenges.}
In everyday scenarios, users commonly handle multiple tasks at once and encounter interruptions. This creates challenges for AR assistance because oftentimes the system does not recognize that the user is goal switching and so responds incorrectly~\cite{DBLP:conf/iui/BonanniLS05}. In these cases, efficacy will be affected, which can be detrimental to the user's trust of system~\cite{DBLP:conf/wirtschaftsinformatik/ZierauE0L20, DBLP:conf/chi/MahmoodFW022}.

\input{tables/experts}
\begin{figure*}
\centering
\begin{tabular}{cc}
  \includegraphics[width=50mm]{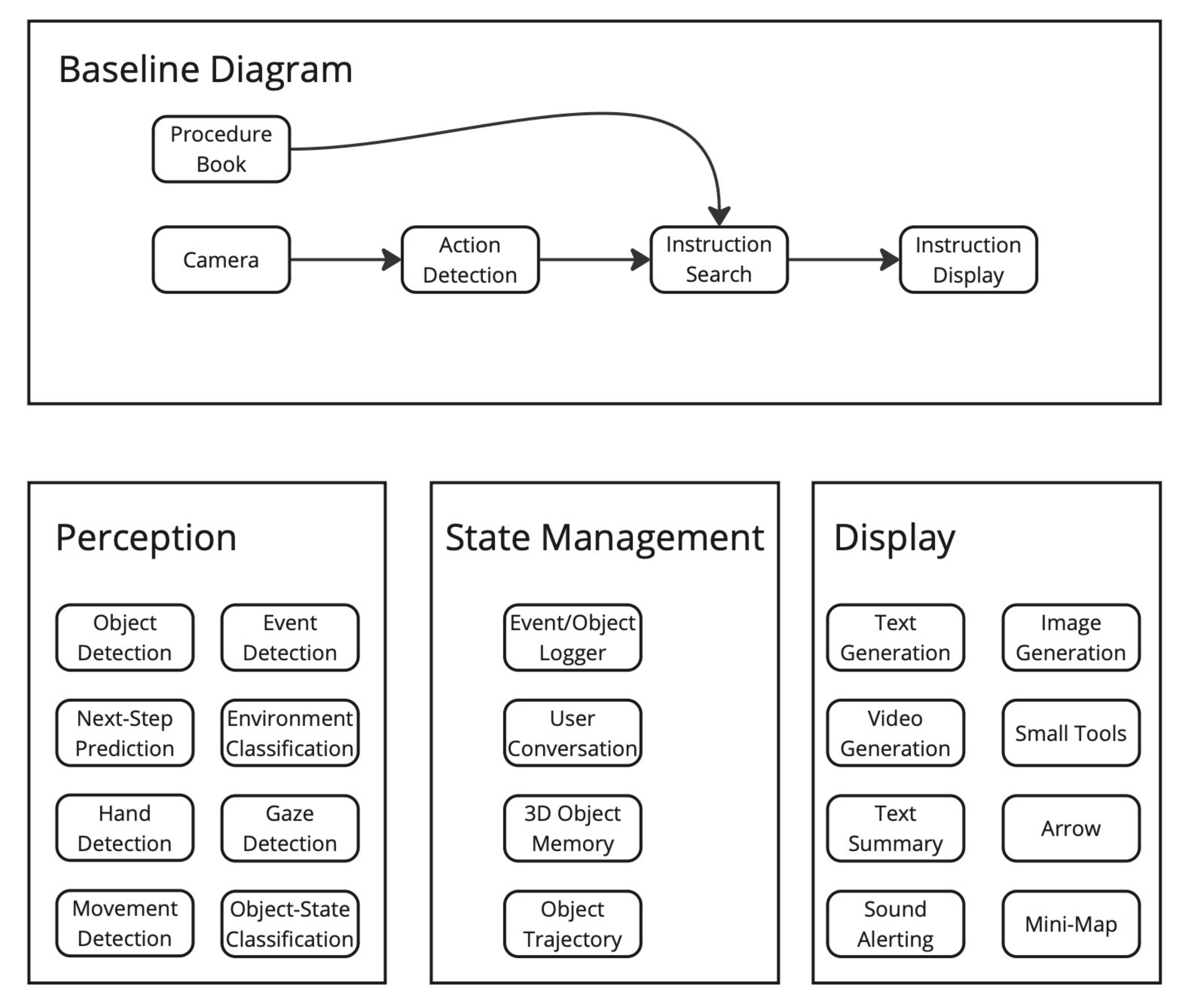} &   \includegraphics[width=75mm]{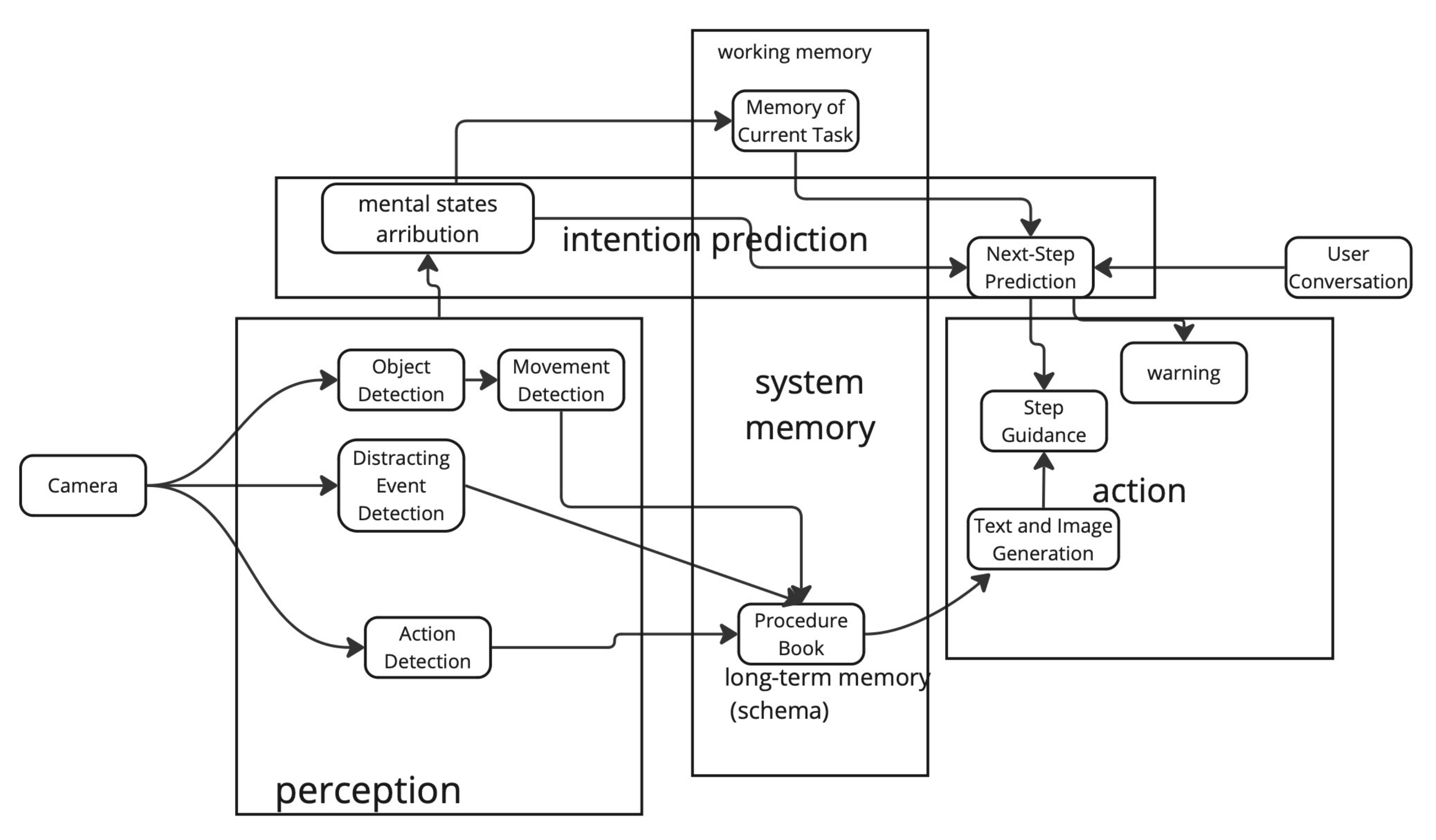}  \\
(a) Initially presented diagram and available modules. & (b) Sample results from Group B. 
\\[6pt]
\end{tabular}
\caption{During the first session (participatory design), experts need to collaborate on creating an ideal assistant framework based on the presented diagram and modules. At the bottom of Figure (a), the experts can find the system components for perception. Figure (b) is a result of the original diagram illustrated by one expert group.}
\label{fig:formative-session1}
\end{figure*}

\subsection{Interview One: Participatory Design} 
To formalize \textbf{how to design a proactive system} capable of determining what to show users for task completion, we presented the known challenges and background knowledge to the experts as described in Section~\ref{sec:codesign:background}. During the presentation, we described the interaction context, explained the capabilities of current AR technology, and clarified any concerns the experts raised. Each group was then asked to discuss: 1) the information necessary for the AR system to act proactively; 2) any necessary system features, methods, or functions; 3) the perspective helpfulness of user modeling; and 4) ways to mitigate known challenges. 

Each group was then moved into their own private discussion room. After a 50-minute open-ended discussion, we provided each group with a list of commonly used tracking, computer perception, contextual understanding, and display technologies and let them select which to use, see Figure~\ref{fig:formative-session1} for reference. Experts were invited to add ``imaginary'' categories or functions to this list if they considered it theoretically useful. Their modified lists were illustrated using Miro~\footnote{https://miro.com}.

\subsection{Interview Two: Adaption of Design Models for AR} 
The second session involved reconvening the same groups of experts for dyadic interviews. Initially, we presented the outcomes of the first session alongside our synthesized framework, seeking confirmation that it accurately reflected their initial ideas. This was followed by an open discussion where the experts delved into the framework's details and made adjustments to further refine it. This session, which lasted approximately one hour for each of the three groups of experts, was essential for finalizing the design framework for the AR assistant. 

\subsection{Data Collection}
Since the interviews were conducted over Zoom, we screen-recorded and transcribed the interviews using Zoom's auto-transcription feature. Two authors independently analyzed the video recordings and transcriptions, coding the findings into insights. The insights were then combined into the following findings based on thematic analysis, and discrepancies were resolved through discussion.

\subsection{Results}

\paragraph{The BDI model may be a good candidate for supporting proactive guidance}\revision{ During the interviews, all three psychology experts (E2, E4, and E6) mentioned that considering \textbf{What the user sees and understands in the surroundings} is important for predicting when guidance should appear (C1). For instance, E4 emphasized, \participantquote{... it is important to model the human's mental space, so we can adjust the AR (assistance's) timing.} All the psychology experts introduced \concept{belief-desire-intention}, describing it as well-established and straightforward, as well as a classic cognitive model for understanding human behavior, intention, and goals. }

When describing ideas to implement the BDI model within the AR context, the expert groups outlined how \textbf{belief} supports the filtering of duplicated or unnecessary assistance and acts as a screening step to narrow the assistance's scope. They further outlined that \textbf{desire} refers to the goals of a given task. In AR, this means the system should model the user's actions and goals (Group B and C). Finally, the expert groups indicated that \textbf{intention} comprises a small step toward the goal and affects the timing and content of the AR assistance (Groups A and C). Together these adaptations of the BDI model help to construct a novel pipeline toward proactive AR assistance.


\paragraph{Determining the user's intention is essential to proactive guidance. } 
\revision{
Group A and Group C first brought up the importance of understanding user intention, which they construed as the \textit{immediate step being undertaken in the context of the guidance}. The group claimed that knowing the intention of the user is beneficial for effectively determining the content of the assistance and its timing. Additionally, when discussing how to design  ``next-step prediction'' in practice, E6 suggested that computer vision models might be able to infer user's intention. However, E5 thought otherwise and commented that the common method of inferring intention using egocentric short-term memory cannot predict intention reliably. All groups agreed that new methods are required to infer user intention.}

\input{tables/formative}

\paragraph{Understanding high-level goals improves transparency and efficiency in task switching.}
\revision{
Groups B and C discussed transparency challenges in human-AI collaboration. Interaction can be improved if information on tasks, objects, and goals is available to both the system and the user simultaneously. On the users' end, this is essential to support multi-tasking with task guidance as users are constantly aware of ``how the system interprets the current situation'' (Group A, B). On the system's end, knowing the user's high-level goals (e.g., task goals) can support multi-tasking effectively and automatically (E1 and E2). Additionally, providing the step-by-step reasoning that leads toward task completion is beneficial for users in that it allows them to maintain trust while collaborating with AI (Group A and C).}

\paragraph{Using the potential of modern LLMs for understanding context, environment, objects, and actions.}
\revision{
E5 has extensive experience in traditional computer vision models and expressed concern that current computer vision models may not be sufficient due to the inaccuracy of action and intent prediction. Even if users' intentions (i.e., immediate goals) can be detected, the predicted intention cannot be used to the fullest extent because these models often lack the ability to understand the user's environment or make accurate decisions based on intent predictions. E1, who has significant experience in LLM development, suggested that multimodal LLMs like GPT-4V could offer a solution because of their advanced reasoning capabilities. Exploring prompting techniques may help to detect context, environment, objects, and actions.}

%% file: tables/experts.tex
\begin{table}[]
\begin{tabular}{|l|l|l|l|}
\hline
Expert & Background                    & Gender & Group \\ \hline
E1     & HCI                           & M      & A     \\ \hline
E2     & Psychology                    & F      & A     \\ \hline
E3     & Computer Vision \& Psychology & F      & B     \\ \hline
E4     & Psychology                    & M      & B     \\ \hline
E5     & HCI                           & M      & C     \\ \hline
E6     & Psychology                    & M      & C     \\ \hline
\end{tabular}

\caption{The table shows the experts' backgrounds in the co-design. We paired one computer science expert with one psychology expert per group. In total, three groups participated in the co-design.}

\label{tab:expert}
\end{table}

%% file: tables/formative.tex
\begin{table*}[t]
\centering
\begin{tabular}{p{4cm}p{10cm}}
\toprule
\textbf{Formative Study 1} & \textbf{Results} \\
\midrule
 & \textbf{\textit{Benefits}} \\
 & Could be automatic. \\
 & Reduces development time and increases efficiency. \\
 & Improves scalability. \\
 & Reduces information overload. \\
 & \textbf{\textit{Challenges}} \\
 & Cross-domain scalability is difficult. \\
 & Detecting user intention is a primary challenge, as errors lead to confusion. \\
 & Adapting the AR instruction to users' active duties is challenging. \\
\midrule
\textbf{Formative Study 2} & \textbf{Results} \\
\midrule
 & The BDI model may be a good candidate for supporting proactive guidance. \\
 & Determining the user's intention is essential to proactive guidance. \\
 & Understanding high-level goals improves transparency and efficiency. \\
 & Using the potential of modern LLMs might offer a better understanding of context, environment, objects, and actions. \\
\bottomrule
\end{tabular}
\caption{\textbf{The table summarizes the main results from two formative studies.}}
\label{tab:formative-results}
\end{table*}

%% file: sections/04_method.tex
\section{Design Requirements}
\label{sec:design}
Based on the findings of the two aforementioned formative studies, \revision{as summarized in Table~\ref{tab:formative-results}}, we propose the following design requirements for consideration in proactive AR assistance.
\begin{enumerate}[start=1,label={[\bfseries D\arabic*]}]
\item \label{dr:bdi} \revision{Proactive AR assistance can be challenging to implement due to difficulty in timing its appearance, updating assistance to fit the user's environment, and understanding the user's goals and actions. The BDI model offers a new opportunity to provide real-time, in situ, updated AR content.}   
\item \label{dr:assistance} \revision{AR assistance should convey appropriate content via an appropriate modality at the right time. It should also support users switching tasks or actively manage task life-cycle (i.e., beginning, pausing, and ending).}
\item \label{dr:transparency} \revision{Assistance should try to be transparent to gain users' trust, feed back the system's reasoning and detection, and provide easily accessible information about current and overall goals in the AR environment.}

\item \label{dr:llm} \revision{LLMs could be used to improve scalability and re-usability. Using LLMs might offer a viable way to analyze complex environments, model user action and goals, track progress, and update assistance content in situ. The result would be a more adaptive, scalable system for various common tasks.}

\end{enumerate}

\section{\systemname~System}
Guided by the design requirements, we present the implementation of our proposed \systemname~system. The goal of the implementation is automatic multimodal AR assistance (e.g., instructions, images, illustrations) with appropriate timing and content that is adaptive to the users' immediate surroundings. Through \systemname, we aim to automatically update content to match the context and environment of the interaction, reducing the need for repetitive instructional information toward task completion.

We first use the BDI model as a blueprint to design a workflow to achieve proactive assistance. Next, we detail the implementations for timing prediction and assistance prediction. Finally, we describe our interface and interaction design while ensuring transparency and interpretability.





\subsection{Implementing the BDI Model for AR Assistance}
\input{tables/bdi}
\textbf{Architecture:} We account for the unique characteristics of AR devices and technologies, such as small field of view, the need for continual real-time environmental mapping, and the blend of physical and digital information. We describe how to apply the BDI model in terms of its components. This approach has been used when applying the BDI model to other fields to support intention and goal analysis~\cite{DBLP:conf/aaai/ZhangOY20, DBLP:conf/aaai/YangLQ21}. We follow a similar approach and implement the system architecture as in Figure~\ref{fig:bdi-system-architecture}. The results of the implementation is also summarized in Table~\ref{tab:bdi-ar-guidance}.

\begin{figure*}
\centering
  \includegraphics[width=\linewidth]{figures/user-model6-compressed.pdf} 
    \caption{The figure is a system overview of the BDI user model. The system processes inputs from the camera's view, dialogue (voice communication between the user and the GPT model), and the historical logger (records of prior assistance). These inputs are sent to different BDI components for analysis and inference using a combination of local models and LLMs to generate proactive guidance and determine the appropriate modality and assistance timing. To ensure assistance appears and disappears at the right time, a task planner LLM generates a step-by-step task plan based on the inferred desire, with multiple checkpoints assigned to each step. These checkpoints are monitored by the action finish detection module, which determines task completion by verifying checkpoint progress. In addition, the system employs an early forecasting module to minimize latency.}
    \label{fig:bdi-system-architecture}
\end{figure*}

\subsubsection{BDI-guided chain-of-thought}
On a high level, the BDI model aligns with the concept of chain-of-thought (CoT)~\cite{wei2022chain} in LLM. CoT is a form of reasoning that allows the LLM to deliver assistance in a structured manner by sequentially following logical steps. By conceptualizing the BDI model as a series of thoughts, the model can systematically produce the appropriate assistance. Each thought in the process is marked with a hashtag, enabling the LLM to decompose complex tasks into manageable steps, thereby enable reasoning functions (e.g., action prediction, task prediction, guidance, etc.) in AR assistance. The following subsections describe how we conceptualize the BDI model.

\subsubsection{Belief}
\label{sec:belief}
Human \concept{belief} is a complex psycho-neural function integrally connected with memory and cognition~\cite{lund1925psychology, porot2021science}. Precise modeling of human belief within the constraints of AR technology is not feasible without access to human neural signals. To approximate the user's belief state within AR constraints, we propose a two-fold method via capturing scene and objects from the AR's visual input and via user action history from task performance. 

The \textbf{scene} provides information on the user's surrounding physical setting, the context of the ongoing task, and changes in their goals and actions. We represent the scene via the label predicted by the image classification model. The label prediction uses an OWL-ViT model~\cite{minderer2022simple}, which is the zero-shot object detection model. The scene detection is implemented with the zero-shot image recognition model CLIP model ~\cite{radford2021learning}.

\textbf{Object} information could be used to locate and filter task-relevant objects in the scene from others. To achieve this, we used two different models for object detection: Detr model to detect objects in the scene in zero-shot~\cite{carion2020end}; and LLaVA model to detect objects that are being held/touched/moved by human hands~\cite{liu2024visual}. We did not use fixed-label set models because they cannot cover the entire case.
We did not use the traditional object detection models in this case because these models are trained to predict a fixed set of labels, limiting generalizability.

\textbf{Action and assistant history} is used to ensure the guidance does not repeat. Due to the nature of linear task guidance, completed steps or instructions should not reappear. In our earlier testing, we noticed that the model prediction may give the same instructions that had appeared previously despite task progression. As a result, we implemented a history log to reduce such repetitions. This history contains user interaction logs, the AR assistance content, descriptions, progress, modalities, and images. 
\subsubsection{Desire}
\label{sec:desire}
This component represents the user's high-level goals, or task goals for the AR system. From cleaning a room, to preparing food, to organizing a shelf, high-level goals are short-term tasks users aim to accomplish. Inspired by recent work that successfully used LLM to understand instructional tasks, \revision{we infer the user's goals using a GPT-4V, which takes the current camera frame as input to predict the high-level goals. Image frames are downsampled to 1 fps and sent to the LLM with a prompt specifying the need to understand ``what the user is doing, at what place''. The resulting label from the GPT-4V contains the task's general description (e.g., moving a table, arranging desk, etc.). }

However, the current LLM does not always predict the goals correctly. Our initial testing revealed 85\% accuracy in predicting the correct task goals in a common household settings. As a result, instead of \systemname~ immediately beginning to instruct the user after detecting their goals, it first asks users to \textbf{confirm} the predicted tasks or goals. This allows the users to begin AR guidance only if they accept \systemname's suggestion of a given task, ensuring error-free task launching.

\subsubsection{Intention}
The results of the formative studies established that the concept of intention from the BDI model could affect the content, timing, and upcoming actions required to complete a task. To predict the user's upcoming actions, we rely on perceptual information (\ref{dr:bdi}), including visual cues and user interactions with objects. We use a combination of localized models with LLM to balance the time cost for timing prediction. As for content prediction, we use customized prompts and CoT coupled with GPT-4V's semantic understanding to determine what type of assistance might be needed.

\subsection{Timing Prediction}
To determine when assistance should appear the system must first detect a user's action and then the corresponding assistance follows. We begin with a step-by-step pipeline to predict when an action will occur. The first naive implementation performs action-forecasting after the previous action is completed. This is achieved by concatenating the last four frames and sending them to GPT-4V model via OpenAI's API at 1 fps. \revision{However, since the model prediction from LLM is not instant, the user must wait for the prediction to display after actions are finished, resulting in their interaction experience being interrupted. To correct this, we use a combination of \textbf{action forecasting} and \textbf{early forecasting} to reduce the interaction latency and provide a seamless experience.} \revision{When the system is running, it continuously executes action forecasting using LLM; meanwhile, parallel early forecasting focuses on detecting action completion. Once detected, cached actions from the continuous action forecasting are immediately retrieved and the assistance is displayed. This way the user no longer has to wait (what was typically about 3 extra seconds) after their action was finished to move forward.}

\subsubsection{Action forecasting}
We propose a multimodal LLM to forecast upcoming user actions. This is challenging due to the vast range of potential future actions, the ambiguous nature of user goals, and the misalignment with the label set. We start with constraining the forecasting process by incorporating the user's high-level goals, thus narrowing down the range of possible actions. We then prompt these actions to the LLM using a search-and-reflect framework consisting of three stages:
\begin{enumerate}
    \item \textbf{Analysis Stage:} The LLM first analyzes the current task goals and corresponding task plan (see Section~\ref{sec:taskplanner}), breaking it down into actionable steps.
    \item \textbf{Prediction Stage:} After analyzing the goals and plans, the LLM determines the upcoming actions. This involves using contextual cues (e.g., physical objects, scene, and the user's action history) and the results from task planner to converge on several probable actions.
    \item \textbf{Reflection Stage:} The LLM further narrows to the single predicted action (or next step) by integrating the objects and tools in the scene. Actions that require missing or unavailable objects are eliminated, ensuring that only viable actions are suggested. This filtering helps refine the prediction further by aligning it with the actual scene context, reducing irrelevant or impossible options.
\end{enumerate}

\subsubsection{Early forecasting with finished action detection}
\label{sec:action_finish}
\revision{Early forecasting prioritizing response time to serve as a flag to retrieve action forecasting cached results. The action finish detection detects a series of checkpoints (see Section~\ref{sec:taskplanner}), or mini-goals within each step. If all checkpoints are reached, the action detection is complete. It is important to reduce the detection noise, such as the user not looking at the task or another person coming into view. Since there are no pre-trained models or large-scale datasets for detecting when an action is finished, we use the zero-shot learning capabilities of the vision-language model and propose an ensemble-based approach to balance latency and effectiveness. } We ensemble the local image captioning model BLIP-2 ~\cite{DBLP:conf/icml/0008LSH23} with the online GPT-4V model. \revision{BLIP-2 model has lower accuracy, and this pipeline double checks its result with the GPT-4V model, which produces more reliable action prediction results based on our initial testings. BLIP-2 model also continuously outputs the prediction of where the user is looking, notifying the AR assistant if the user is distracted and filtering out noise.  }

\subsection{Dynamic Content Generation}
The content of AR assistance comes in different forms and via different modalities; inspired by the AR designers in the first formative study, we implemented text, image, sound, and tools for \systemname.  Each has different functions and use cases relative to scene context and user actions. 
\begin{enumerate}
    \item \revision{\textbf{Text:} We use white text on a black, transparent container to ensure readability. The text primarily contains general instructions (task names, titles, etc.), interface information, and step-by-step guidance. All text is dynamically generated from either the LLM's response or sub-steps from the task planner module, see \revision{Figure} for details.}
    \item \revision{\textbf{Image:} Images are generated in situ using DALL-E 3  to depict actions and objects, see examples in Figure~\ref{fig:satori-naive-figure-comparison}. For more complex actions, we employ multiple images. See appendix for the implementation details. }
    \item \revision{ \textbf{Sound:} We use the headset's text-to-speech module for: 1) answering user's spoken responses; 2) reading instructions aloud; and 3) confirming task completion.}
    \item \revision{\textbf{Tools:} We implemented three example tools as a demonstration. Additional tools could be added to the pipeline if needed. \textit{A voice-assistant} that is triggered by the keyword ``Hello Tori'' will listen and respond to voice input and can be used to command system actions with words such as ``yes'' or ``cancel''. If the system thinks the task step requires time counting (e.g., boiling water, microwaving, grinding powders), a \textit{timer} automatically appears. This is achieved by comparing the objects in the scene with objects needed for the current step in the task using the LLM's reasoning ability. \textbf{Object indicator} locates the ``objects of interest'' in the current step. This is done through the object detection methods described in the earlier Section~\ref{sec:belief}. }
\end{enumerate}





\revision{\subsection{Inferring Modality}
We use a GPT-4V to determine the modality using a set of rules in a prompt. The rules map a relationship between the intention and the current step to the corresponding modality. Based on the suggestions from the second formative study, we implemented four rules and their corresponding modality mapping: 1) for intention or steps involving a tool or interaction with materials the LLM returns an image; and 2) if the action is time relevant, the LLM gives a sequence of images; 3) if time counting is needed, the LLM shows the timer tool; and 4) if the step is challenging, the LLM asks for audio feedback. These rules are not mutually exclusive and could generate a combination if multiple conditions are met. 
}

\subsection{Task Planner for Checkpoints}
\label{sec:taskplanner}
\revision{This component first retrieves the most compatible task from a task database once the user's goal is set (see Sections \ref{sec:belief} and \ref{sec:design}). It then provides detailed step-by-step instructions and layout \textbf{checkpoints or sub-steps} for the AR assistance. Each checkpoint is an actionable sub-step to reach the current step completion. The benefit is twofold: 1) It increases system transparency and builds trust for users as each checkpoint is explicitly listed on the AR interface, and 2) it decomposes the step prediction into smaller milestones for the system, increasing overall prediction validity. }

\begin{figure*}[h]
    \centering

    \begin{subfigure}{0.2\textwidth}
        \centering
        \includegraphics[width=\textwidth]{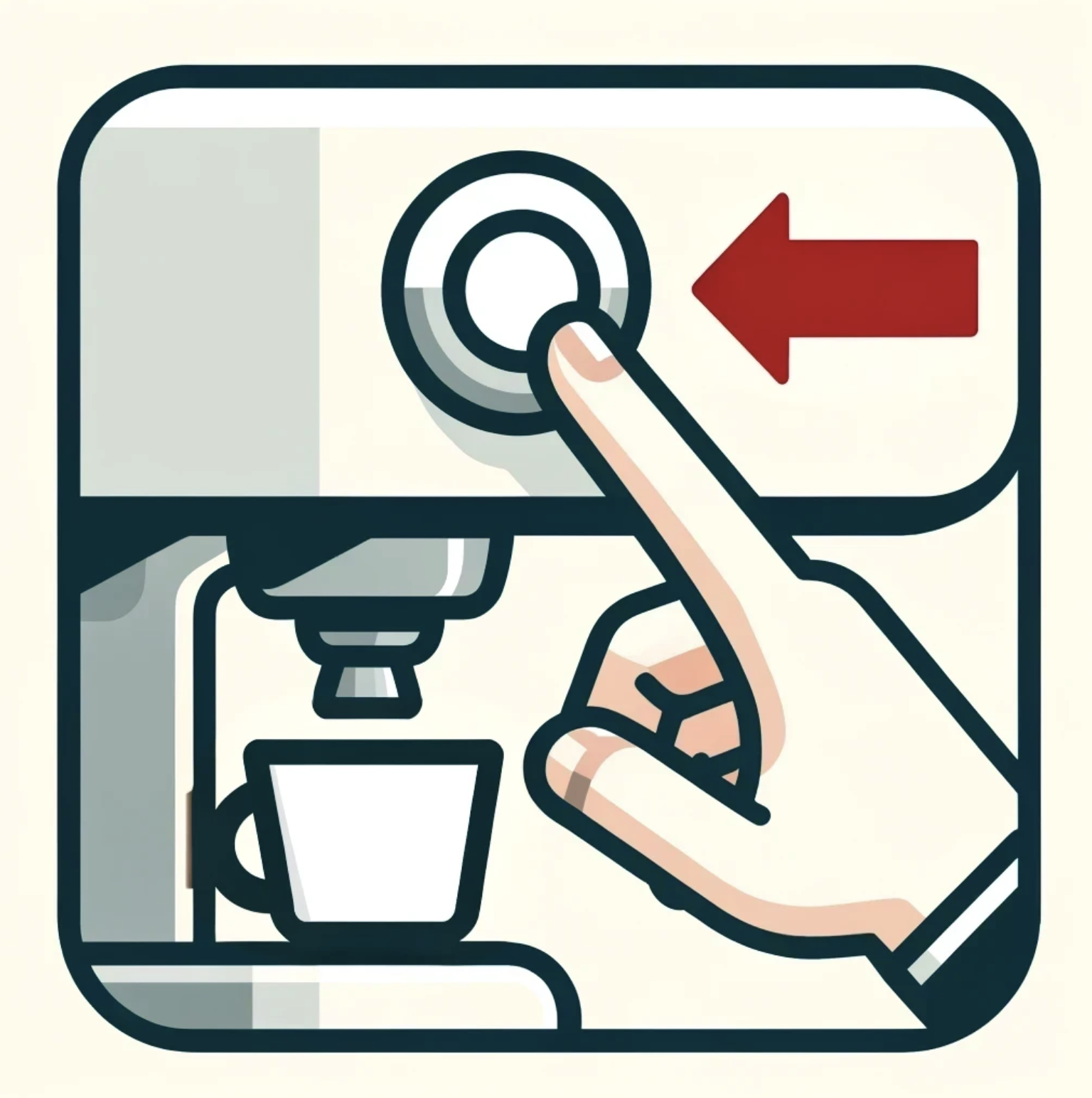}
        \caption{\systemname}
        \label{fig:comparison-a}
    \end{subfigure}\hfill
    \begin{subfigure}{0.2\textwidth}
        \centering
        \includegraphics[width=\textwidth]{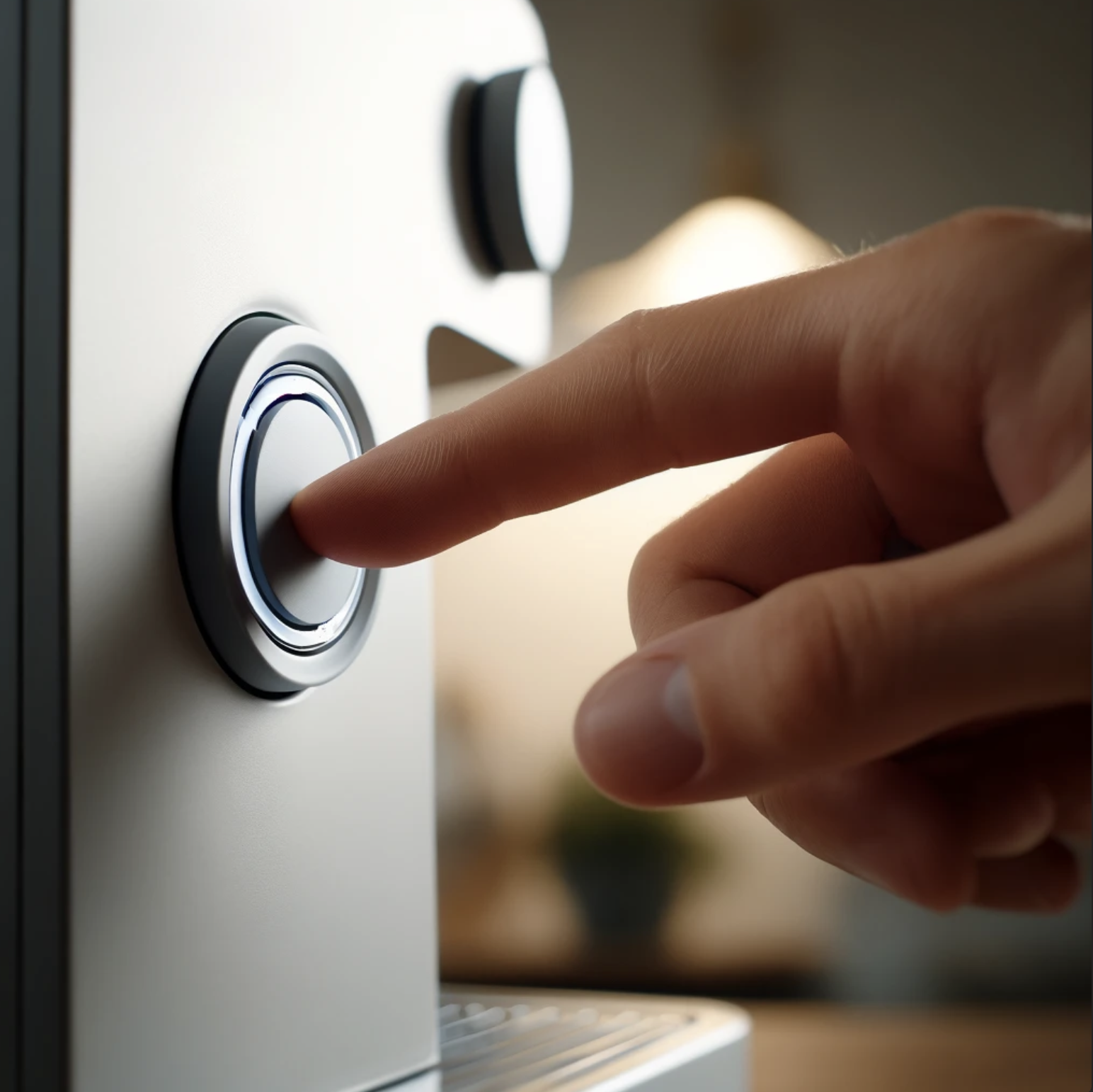}
        \caption{Naive}
        \label{fig:comparison-b}
    \end{subfigure}\hfill
    \begin{subfigure}{0.2\textwidth}
        \centering
        \includegraphics[width=\textwidth]{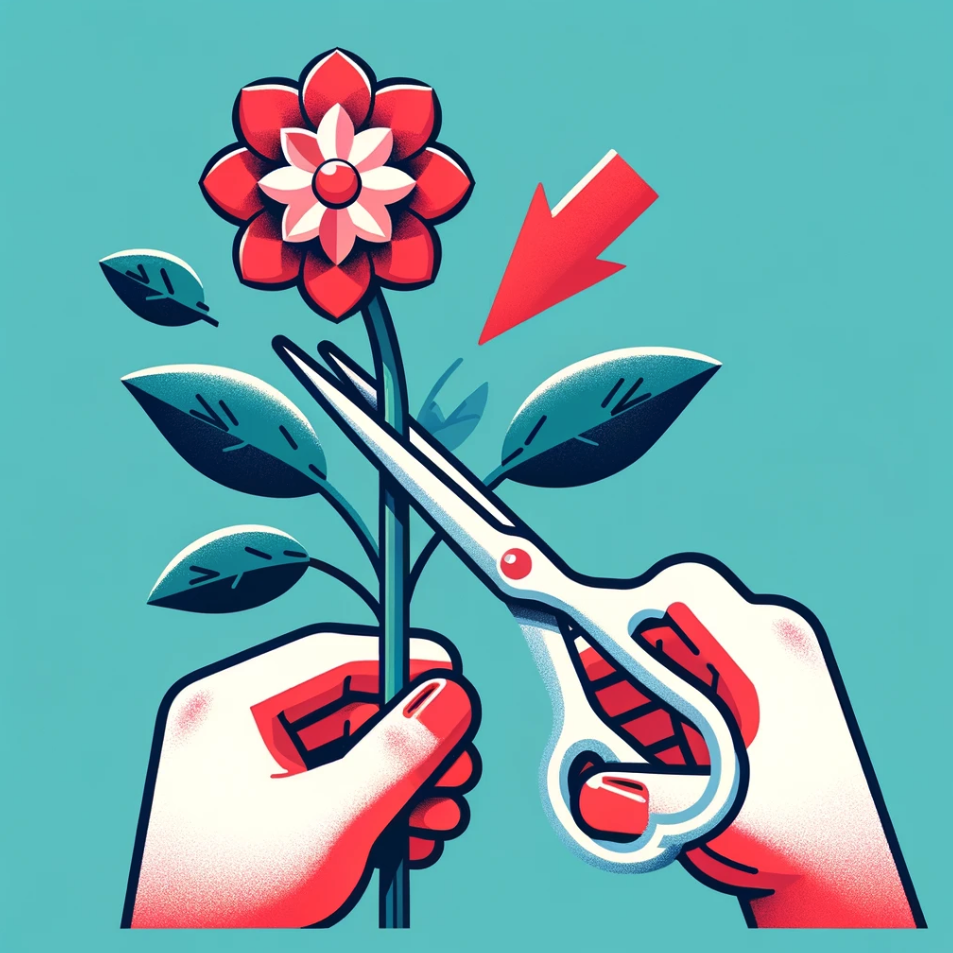}
        \caption{\systemname}
        \label{fig:comparison-c}
    \end{subfigure}\hfill
    \begin{subfigure}{0.2\textwidth}
        \centering
        \includegraphics[width=\textwidth]{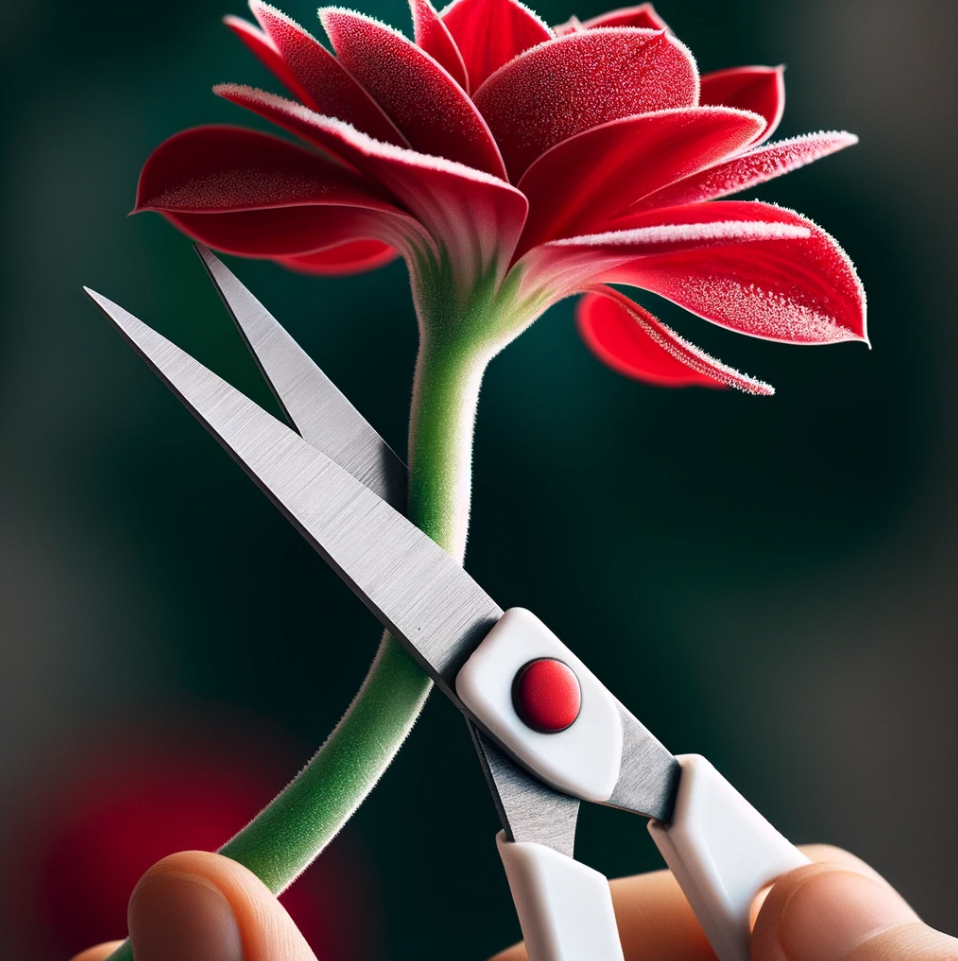}
        \caption{Naive}
        \label{fig:comparison-d}
    \end{subfigure}
    \caption{Comparison of the naively generated images from the GPT model (i.e., Naive) with our proposed prompts (i.e., Satori). (a) ``\textit{One hand presses a white button on a white espresso machine. A large red arrow points to the button. No background, in the style of flat, instructional illustrations. Accurate, concise, comfortable color style.}'' (b) ``\textit{One hand presses a white button on a white espresso machine.}'' (c) ``\textit{Cut stem of a red flower up from bottom, with white scissors at 45 degrees. One big red arrow pointing to bottom of the flower stem. In the style of flat, instructional illustrations. No background. Accurate, concise, comfortable color style.}'' (d) ``\textit{Cut stem of a red flower up from bottom with white scissors at 45 degrees.}''}
    \label{fig:satori-naive-figure-comparison}
\end{figure*}

\begin{figure*}
\centering
\begin{tabular}{cc}
  \includegraphics[width=65mm]
  {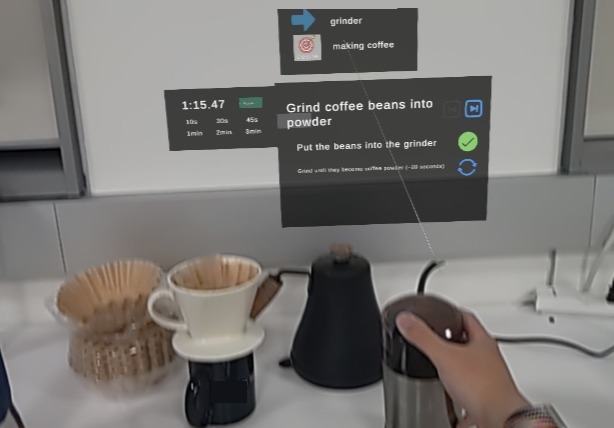} &   \includegraphics[width=65mm]{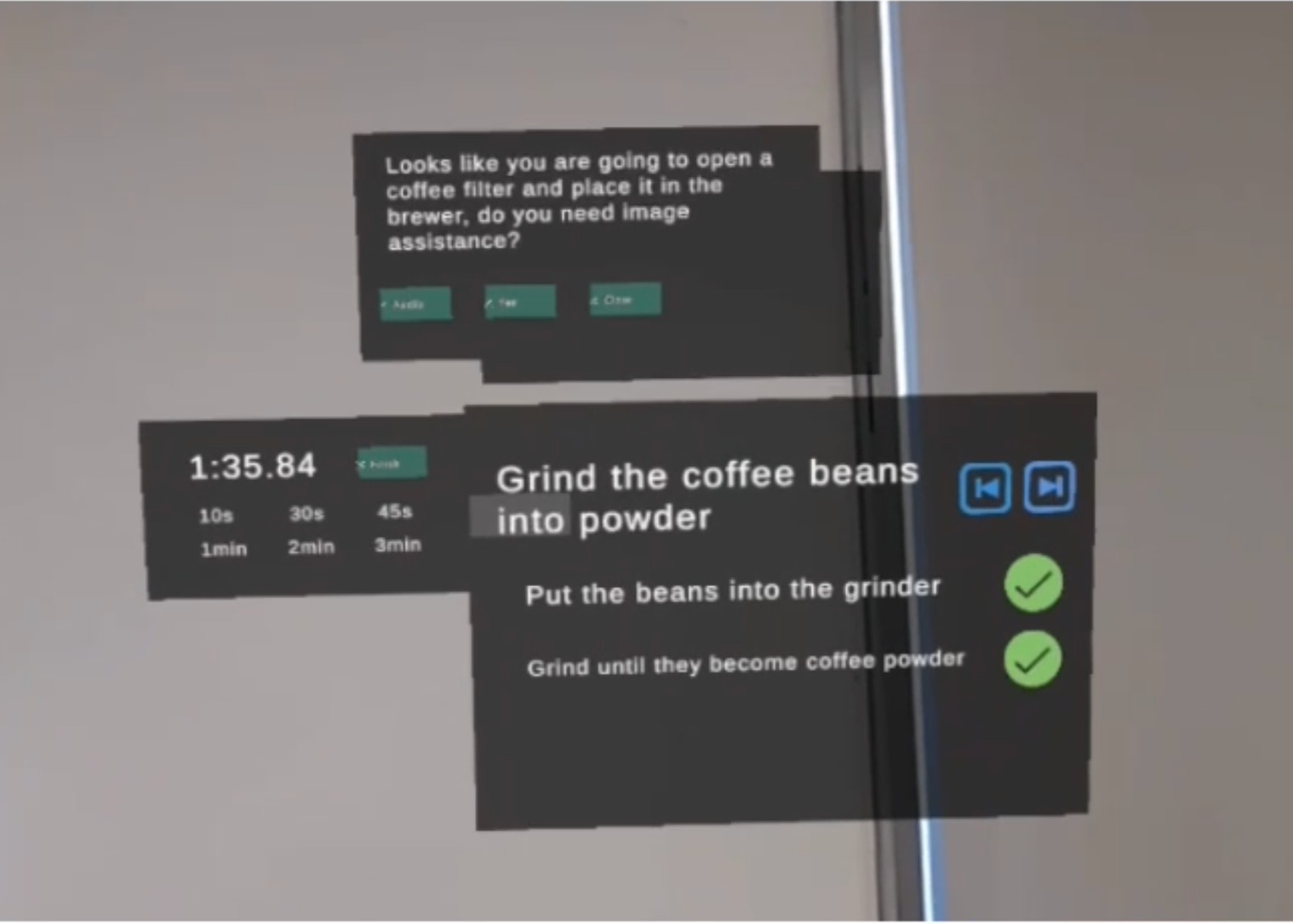}  \\
Interface displays (a) user's upcoming action (Desire) and goal & (b) the confirmation page on top. 
\\[6pt]
\end{tabular}
\caption{ (a) In this example, the user is grinding the coffee beans. The interface shows the task goal as ``Making Coffee'' and the upcoming action or step as ``Grind coffee beans into powder.'' The action checkpoints marked with green checks indicate the number of sub-steps that are completed. The action checkpoints marked with a blue circle indicate the number of sub-steps that are in progress. Once all sub-steps are checked, the current step is considered complete; and (b) A task assistance confirmation appears when the system detects step completion. The confirmation prompts the user, asking if they are about to use a coffee filter and whether they need assistance.}
\label{fig:interface-ui-demo}
\end{figure*}

\subsection{Interface and Interaction Design}
\label{sec:interface}
Figure~\ref{fig:interface-ui-demo} displays the interface with assistance, including active task (e.g., \textit{Make Coffee} in the example), text instructions, images, and tools such as the object indicator (e.g., \textit{coffee grinder}) or timer. Aligning with design requirement to remain transparent~\ref{dr:transparency}, \systemname's interface shows how the system tracks the user's task, their progress, and objects of interest. For example, the object indicator not only shows the object that the user needs to interact with but also points to the object's physical location relative to the user. 

The voice feature is also supported to let users communicate with \systemname~hands-free. The Voice interface is activated if the user calls out the activation phrase or during any confirming stage such as goals or step confirmation. This allows the users to quickly express their intentions without interrupting the tasks on hand.

\subsubsection{Human-AI interaction design}
In the early testing, we found that users could become overwhelmed if the predicted action changed abruptly. \revision{This is because no existing systems can perfectly predict user action, and not every action is meaningful for the task (i.e., behavioral noise).} \revision{Moreover, due to the nature of step-by-step guidance, prediction errors tend to accumulate across steps, and, without human correction, errors in earlier steps may propagate to later steps.} Therefore, we opted to use a confirmation panel to determine whether the system's task or action prediction matched the user's intention, as shown in \revision{ Figure \ref{fig:interface-ui-demo}. For example, in the coffee-making task, if AR assistance failed to detect that the coffee beans had been grounded, it might continuously prompt the user to grind the beans. With \systemname, the system prompts a confirmation page, waiting for the user to confirm action completion. No additional information will appear to the user before they confirm the step with either the pinch button or voice. Similarly, when a new task or step is detected, the confirmation page displays, and users decide whether it matches their needs or the current step, as shown in Figure ~\ref{fig:interface-ui-demo}(b).}

%% file: tables/bdi.tex
\begin{table*}[ht]
\centering
\begin{tabular}{|l|p{2.5cm}|p{3cm}|p{3cm}|p{3cm}|}
\hline
\textbf{BDI Comp.} & \textbf{Definition} & \textbf{AR Guid. Comp.} & \textbf{Inference Method} & \textbf{Example Usage} \\ \hline
\multirow{3}{*}{Belief} 
& \multirow{3}{=}{Representation of the world}  & Scene understanding & OWL-ViT for zero-shot scene classification & Minimizing distractions caused by head movement \\ \cline{3-5} 
& & Task-relevant object detection & DETR for object detection, verified by LLM & Locating objects to improve task efficiency \\ \cline{3-5} 
& & User action history & Logged by an in-memory logger and inferred by LLM & Preventing repeated instructions for completed steps \\ \hline

Desire & Goals or objectives & High-level task goal & LLM-based scene analysis with user confirmation & Assisting task transitions with accurate goal identification \\ \hline

\multirow{2}{*}{Intention} 
& \multirow{2}{=}{commitments that are actively pursued to achieve goals}  & Next intended action & GPT-based inference with CoT reasoning & Providing step-by-step guidance for upcoming actions \\ \cline{3-5} 
& & Timing of next action & Checkpoint-based early forecasting & Reducing latency in delivering next guidance \\ \hline
\end{tabular}
\caption{This table illustrates how three components--- Belief, Desire, and Intention-- in the BDI model are adapted for AR task guidance. BDI Comp. refers to the BDI components and AR Guid. Comp. refers to the AR system's task guidance components. \textbf{Belief} is represented through scene understanding, task-relevant object detection, and user action history to minimize distractions, locate objects, and avoid repeated instructions. \textbf{Desire} captures the user’s high-level task goals, inferred through LLM-based scene analysis and confirmed by the user to ensure accuracy. \textbf{Intention} includes predicting the following intended action using GPT-based inference with chain-of-thought reasoning and determining the timing of next actions with checkpoint-based early forecasting.}

\label{tab:bdi-ar-guidance}
\end{table*}

%% file: sections/05_user.tex
\section{Evaluation}
We evaluated \systemname~prototype through an open-ended exploratory study, focusing on the following research questions:
\begin{enumerate}
    \item \label{rq:1} Can \systemname~provide the correct assistant content at the right \concept{timing}?
    \item \label{rq:2} Can \systemname~provide \concept{comprehensible and effective} guidance?
    \item \label{rq:3} How does our system's guidance compare to that of the professional AR experts? 
\end{enumerate}


\subsubsection{Tasks}
\begin{figure*}[t]
    \centering

    \begin{subfigure}{0.24\textwidth}
        \centering
        \includegraphics[width=\textwidth]{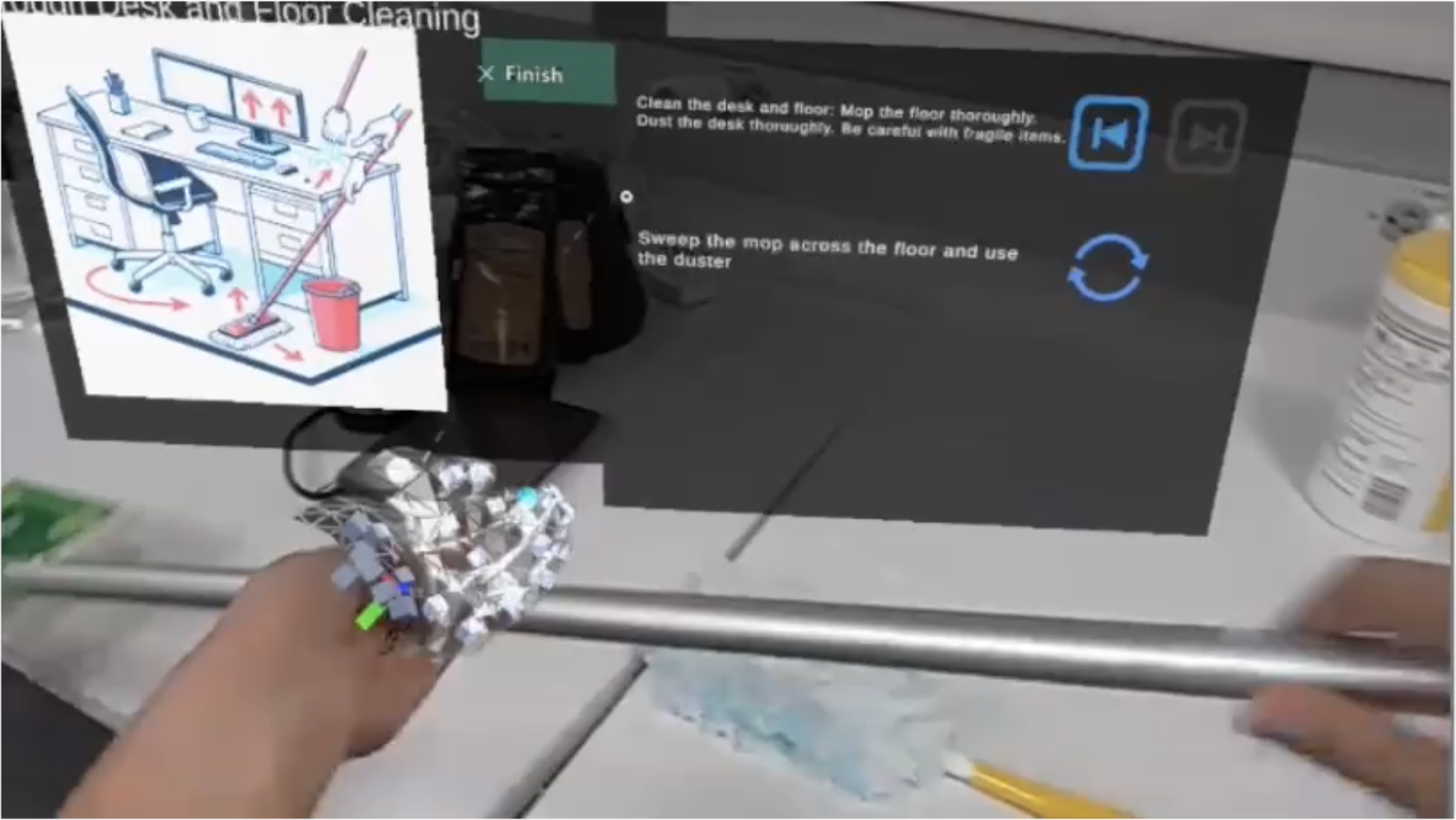}
        \caption{\systemname: clean room}
        \label{fig:study1}
    \end{subfigure}\hfill
    \begin{subfigure}{0.24\textwidth}
        \centering
        \includegraphics[width=\textwidth]{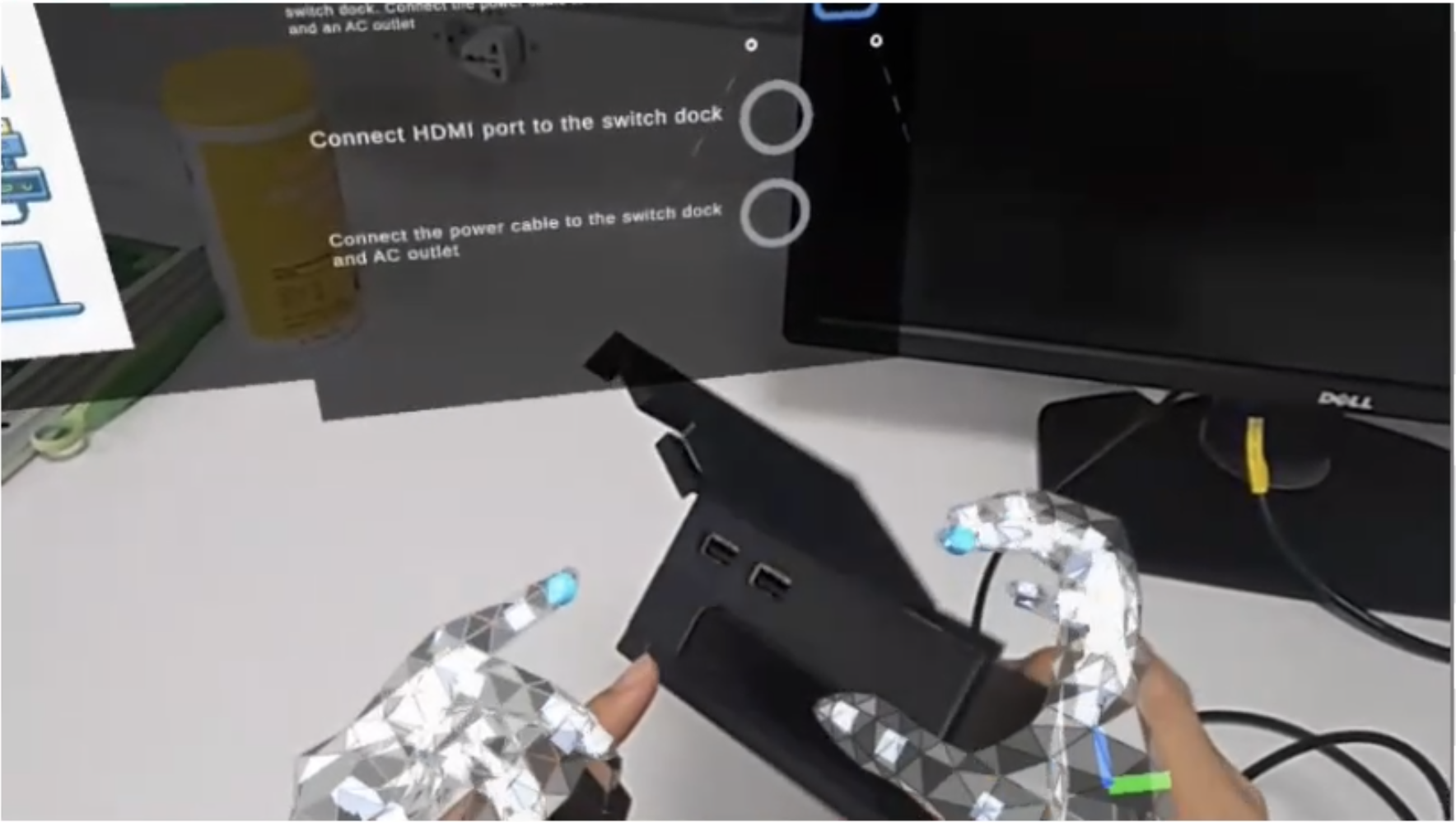}
        \caption{\systemname: connect Nintendo}
        \label{fig:study2}
    \end{subfigure}\hfill
    \begin{subfigure}{0.24\textwidth}
        \centering
        \includegraphics[width=\textwidth]{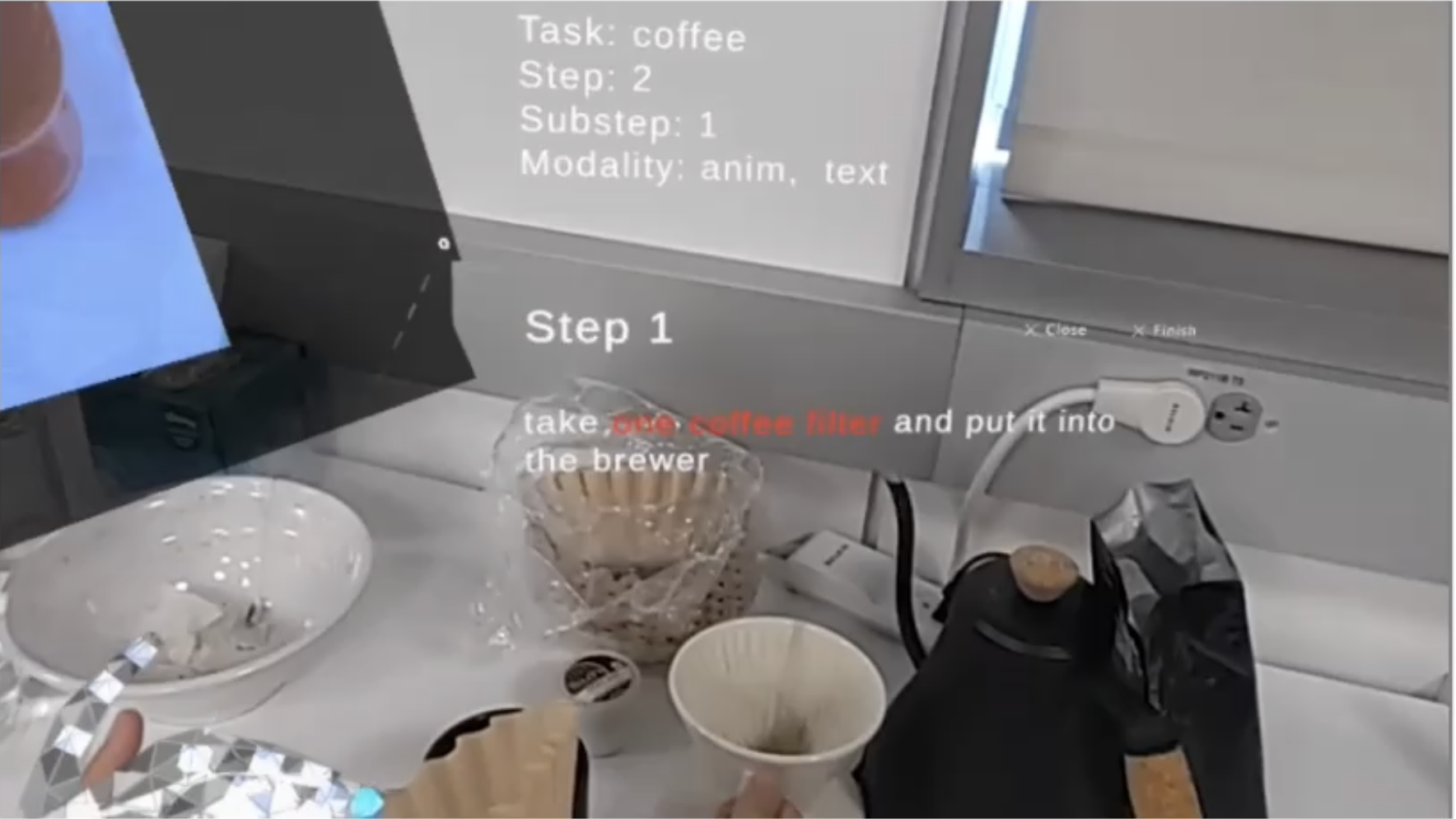}
        \caption{\baseline: make coffee}
        \label{fig:study3}
    \end{subfigure}\hfill
    \begin{subfigure}{0.24\textwidth}
        \centering
        \includegraphics[width=\textwidth]{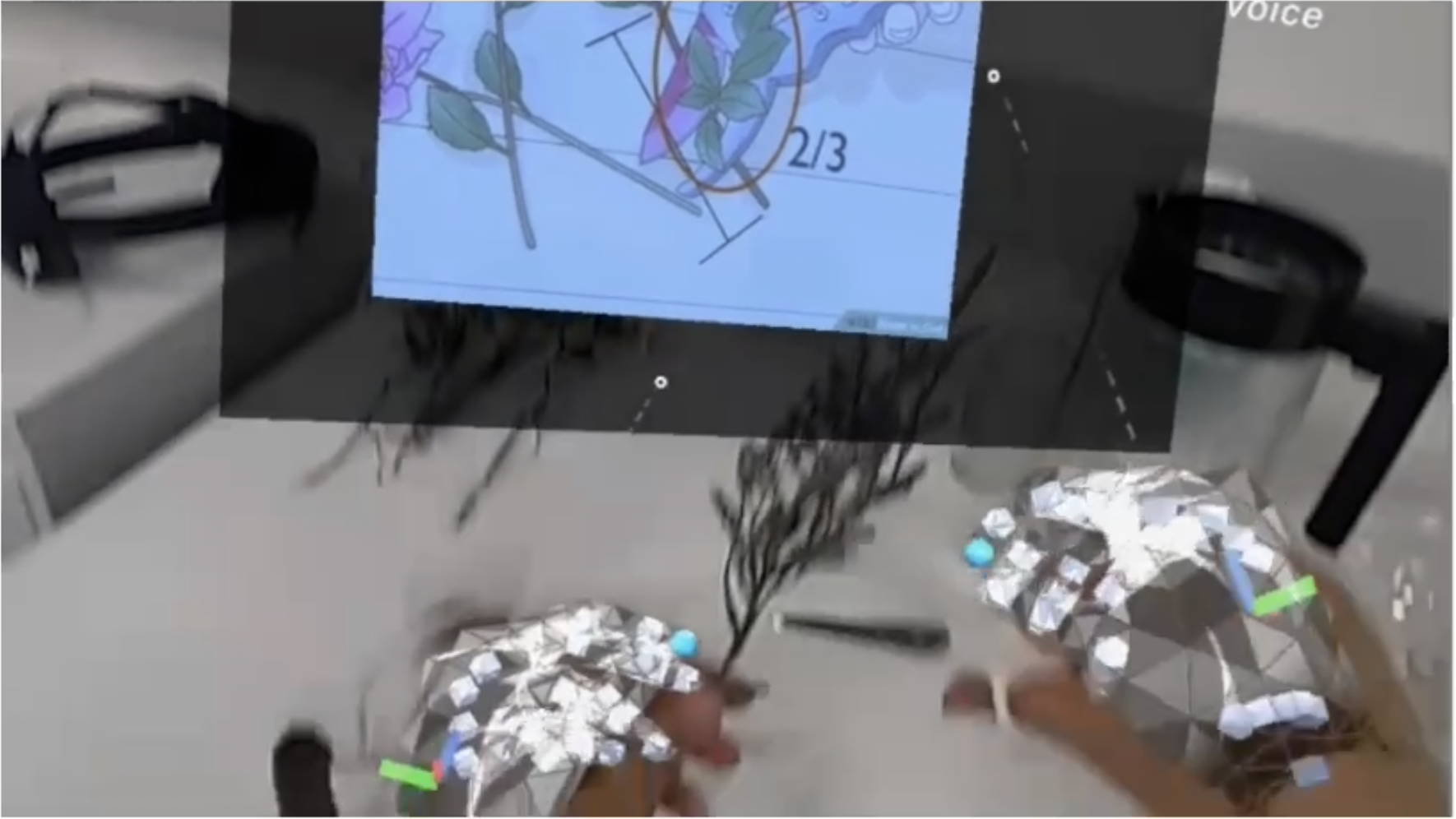}
        \caption{\baseline: arrange flowers}
        \label{fig:study4}
    \end{subfigure}
    \caption{Evaluation tasks using either Satori or a Wizard-of-Oz baseline. (a) The participant is assembling a mop during the room-cleaning task; and (b) The participant is connecting an HDMI cable to a Nintendo Switch dock during the connecting Nintendo Switch task; and (c) The participant is preparing a filter during the coffee-making task; and (d) The participant is trimming flower stems during the flower-arranging task.}
    \label{fig:study_procedure}
\end{figure*}

For our main tasks, we chose four everyday tasks that are comparable in difficulty but different in their goals and required skills, as shown in Figure \ref{fig:study_procedure}. The four tasks were initially sampled from WikiHow~\footnote{https://www.wikihow.com/} and were subsequently rewritten to ensure a consistent task load. \revision{Each task asked for specific sequencing and approach, minimizing users' ability to jump ahead of the instructions using prior knowledge. The task orders were pre-determined and counter-balanced for all 16 participants to avoid the ordering effect. The tasks were as follows:}

\begin{enumerate}
    \item \textit{Arranging Flowers:} Participants arranged a variety of flowers in a vase, testing the system's ability to provide accurate and aesthetic guidance.
    \item \textit{Connecting Nintendo Switch:} This task involved setting up a Nintendo Switch with a monitor, evaluating the system's technical guidance\revision{,} and troubleshooting support.
    \item \textit{Room Cleaning:} Participants assembled a mop and a duster, and cleaned a desk and the floor; the AR assistant suggested assembly instructions and a cleaning strategy.
    \item \textit{Making Coffee:} This task required making coffee using the pour-over method, with the AR assistant providing instructions on tool usage and pouring techniques.
\end{enumerate}

\subsection{Conditions}
Participants were presented with two conditions, Wizard-of-OZ (\baseline) and \systemname. The tasks (indexed as 1, 2, 3, and 4) and conditions were presented in a counterbalanced order to mitigate the learning and other sequencing effects.

\subsection{Participants}
A total of 16 participants (P01-P16, 11 male, 5 female) were recruited via a university email group and flyer. The average age was 23.8 with the maximum age at 27 and the minimum age at 21. Ten of the 16 participants had AR experience prior to the study. Each participant was compensated with a \$30 gift card for their participation. Information on general wellness was collected from participants both before and after the study, and no motion sickness was observed following the study. 

\subsection{Apparatus}
We used a Microsoft HoloLens 2 headset as the AR device for the study. Participants used the \systemname~system or \baseline~ system described earlier while performing the tasks. The headset connects to a server with a Nvidia 3090 graphics card to fetch real-time results. 

\subsection{Procedure}
The study began with a brief tutorial introducing participants to the interface of the two AR systems. Afterward, participants were assigned four everyday tasks. They started with either the \baseline~ system or \systemname~ system before alternating to the other condition. After completion of each task, participants evaluated their experience using a usability scale and assessed their cognitive load using the NASA Task Load Index (NASA TLX). We also conducted a brief recorded interview, asking participants about the advantages, disadvantages, usefulness, and timeliness of the two systems. The experiments were supervised by the Institutional Review Board (IRB), and all task sessions were video-recorded. These recordings were securely stored on an internal server that is inaccessible from outside the university. Participants provided consent, and their personal identity was strictly protected. We collected data on participants' well-being both before and after the experiment and observed no significant adverse effects. The duration of the entire study was two hours on average. All participants completed the four tasks using both systems.

\subsection{Data Collection}
\revision{We used the following metrics to measure the users' perspective on how \systemname's content and timing compared to the AR designer's version. Since content is automatically generated, we measured comprehensibility, helpfulness, and overall cognitive load to assess whether our system is capable of generating similar content utility without overwhelming the users.}

\subsubsection{User-rated scale}
For RQ\ref{rq:1} and RQ\ref{rq:2}, we opted to use a seven-point Likert scale  (similar to Lewis et al.,~\cite{lewis1995ibm}), ranging from ``strongly disagree'' to ``strongly agree'' to measure the timeliness, ease of use, effectiveness, comprehensibility, and helpfulness of the AR assistance. Eleven questions were asked in total. For the complete set of 11 questions, see Table~\ref{tab:sus}. We computed the mean and confidence intervals for each question using the bootstrapping method. Specifically, 1,000 bootstrap samples were generated from the original data set for computation with 95\% confidence intervals for the estimation of the uncertainty around the mean.

\subsubsection{NASA Task Load Index}
We used the raw 100-point NASA TLX~\cite{hart2006nasa} form to measure the cognitive load with the six subcategories. Mean and confidence intervals were calculated for the sum of all ratings and each of the subcategories using the bootstrapping technique. 1,000 bootstrap samples were drawn from the original dataset with 95\% confidence intervals to measure the uncertainty surrounding the mean.

\subsubsection{One-Sided Wilcoxon Signed-Rank Test}
A one-sided Wilcoxon signed-rank test was used to determine whether the user-rated scale and the TLX ratings are significant. The goal was to test whether \systemname~ performed similarly to the AR assistance designed by professionals in AR; however, simply verifying that there is no significant difference between them does not ensure the two conditions are similar. Instead, we aimed to test whether~\systemname~was no worse than the \baseline~by a predefined margin $\Delta$~\cite{georgiev2019statistical, lesaffre2008superiority, laster2003non}. 

The test defines $D_i = X_{Ai} - X_{Bi}$ as the difference between the scores for each participant $i$ under Conditions $S$ (\systemname~) and W (\baseline~), respectively. The adjusted difference accounting for the margin is given by $D_i' = D_i - \Delta = X_{Ai} - X_{Bi} - \Delta$. The hypotheses for this non-inferiority test are:
\[
H_0: \text{median}(D') > 0 \quad \text{(A is worse than B by more than \(\Delta\))},
\]
\[
H_1: \text{median}(D') \leq 0 \quad \text{(A is no worse than B by at most \(\Delta\))}.
\]

Similar to the vanilla Wilcoxon signed-rank test, this procedure involves ranking the absolute adjusted differences $|D_i'|$, calculating the sum of ranks for positive ($W^+$) and negative ($W^-$) differences, and using the test statistic $W = \min(W^+, W^-)$ to compute a one-sided p-value. This p-value indicates whether we can reject $H_0$ in favor of $H_1$. We chose the margin value $\Delta_{TLX}=2.5$ for NASA TLX and $\Delta_{us}$ for the usability scale as they represent half of the rating interval. 

\revision{
\subsection{System Evaluation Preparation}
We used the GTEA~\cite{li2015delving}, EgoTaskQA~\cite{jia2022egotaskqa}, study recordings, and our dataset to evaluate Satori. The GTEA dataset contains egocentric videos of participants performing daily life tasks, and the EgoTaskQA dataset contains questions about humans' beliefs in the world and the model's understanding of humans' beliefs. We used the GTEA dataset with 71 labels and leave-one-subject-out cross validation. Since the EgoTaskQA dataset has large amount of data in the test set, we sampled 200 data points for the evaluation. We use the indirect split, which has the more complicated relationship between the actions and the questions. The user study recordings consist of 14 participants who performed the four tasks described in this section. Two participants' recordings were lost due to data corruption. In addition, we added 4 more sets of the four tasks (totaling 16 videos) as our dataset for evaluation. GTEA, EgoTaskQA, and our dataset are used to evaluate the BDI model output, and user study recordings are used to evaluate modality and guidance timing.}

%% file: sections/05x_result.tex
\section{Results}
\subsection{System Evaluation}
We evaluated Satori's module-level performance on the GTEA dataset and video dataset we recorded from the empirical study and testings. For \textit{desire} task prediction, Satori achieved a balanced accuracy of 100$\%$ on GTEA dataset and our dataset (Table~\ref{tab:desire-prediction}). Satori achieved~$66.50\%$ in \textit{belief} inference, matching the state-of-the-art HCRN model~\cite{DBLP:journals/ijcv/LeLVT21} on EgoTaskQA dataset~$69.53\%$ (Table~\ref{tab:belief-scene-prediction}).
The results on intention forecasting (timing and intention) revealed a $78.38\%$ precision to predict user actions (Table~\ref{tab:intention-prediction}). For modality prediction, Satori reached an average of 75.12$\%$ recall in deciding the modality that matches the WoZ designed by AR experts (Table~\ref{tab:modality-prediction}). We discuss the implications of these results in the discussion session. 

\input{tables/desire}

\input{tables/belief}
\input{tables/intention}
\input{tables/modality}

\subsection{Usability Scale}
\label{sec:sus_result}
\begin{figure*}[t]
\includegraphics[width=0.99\linewidth]{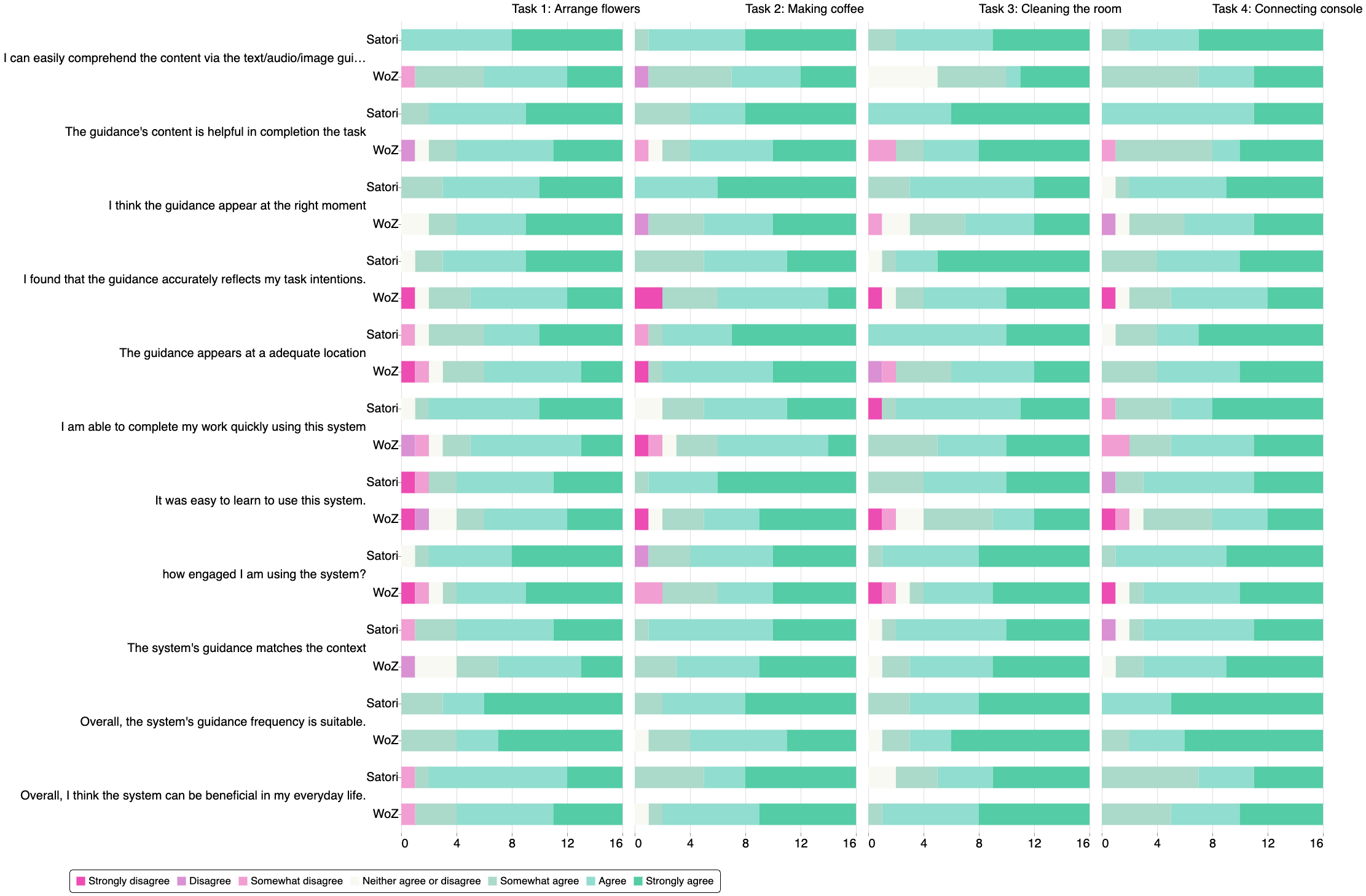}
\caption{Color-coded seven-point Likert scale ratings are shown in the figure for the twelve-participant study. The figure compares the responses for Satori and WoZ systems across four tasks: Arranging Flowers, Making Coffee, Cleaning the Room, and Connecting a Console. Each bar represents the distribution of responses for a specific usability question, highlighting differences in user satisfaction, comprehensibility, and task support provided by both systems.}
\centering
\label{fig:sus}
\end{figure*}

We present the participants' raw scale data across the different tasks in Figure~\ref{fig:sus} and processed statistics in Table~\ref{tab:sus}. We found that there was no significant difference between most of the \systemname~ and the~\baseline~ conditions, suggesting that \systemname's overall performance matched to the wizard-of-oz designed by AR experts ($p_{non\_inferiority} < 0.05$). (e.g., Q1: $p = 0.099$, Q2: $p = 0.094$, Q3: $p = 0.090$, Q6: $p = 0.273$). However, non-inferiority tests demonstrated that \systemname~ was not worse than the~\baseline ~condition (e.g., Q1: $p = 0.001$, Q2: $p = 0.000$, Q6: $p = 0.001$) with a margin of $\delta = 0.5$. 

\subsubsection{Content.} \revision{ \systemname's adaptive AR content provide similar comprehensibility} ($p = 0.099$, non-inferiority $p = 0.001$) and helpfulness ($p=0.094$ and $p_{non\_inferiority} = 0.001$) to complete a guidance task compared to the baseline. \revision{Dynamic assistance almost matches with pre-designed assistance ($p=0.357$, non-inferiority $p = 0.001$).} This is in line with later interview results, where a majority (12/16) believed that \systemname~was able to provide assistance that appropriately matched the context of their tasks.   \systemname's image content is well-received, for example, P1 said that \participantquote{the picture [of the second one] is very nice and it looks good.} Images in the~\baseline~ are also useful, as P8 remarked that \participantquote{Guidance as a whole (text, images, and animations) was very helpful. Whereas, text alone as shown in the image lacks information.}

\subsubsection{Timing.}
\systemname~provides timely guidance to users (Q3: $p=0.090$ and $p_{non\_inferiority} = 0.001$) with appropriate frequency (Q10: $p=0.156$ and $p_{non\_inferiority} = 0.002$). In fact, participants describe the experience as impressive (P16) and can display assistance in need (P3). Although occasional network latency has been reported (P4, P6), they comment that the overall experience was ``not bad''(P6) and ``...sometimes delayed, but I think it's like, it's okay..'' (P4). 

\subsubsection{Effectiveness.}
\label{sec:result:engage}
We found that \systemname~ performs better than the baseline in inferring intention (Q4: $p < 0.05$) and at appearing locations (Q5: $p< 0.05$). Most participants rated between ``agree'' to ``strongly agree'' that AR assistance appears at proper locations in space in both~\systemname~($\Bar{x} = 6.48$) and the baseline~($\Bar{x} = 5.95$). In general, participants felt positive regarding \systemname's assistance effectiveness. P3 stated, \participantquote{I liked that it combines the various modalities of text, audio, and image to generate guidance, I believe that was helpful on multiple occasions where I might have been uncertain with only a single modality.} P14 commented, \participantquote{The guidance helps me a lot, especially in coffee making. It provides me with very detailed instructions including time, and amount of coffee beans I need. I would have to google it if I don't have the guidance.}  P8 noted that \participantquote{For task like arranging the flower vase, the intricate details like trim the leaves, cutting the stem at 45 degree etc. are very necessary details that I might not have performed on my own.} 

In terms of system's learnability (Q7: $p = 0.179$ and $p_{non\_inferiority} = 0.001$) and engagement (Q8: $p = 0.145$ and $p_{non\_inferiority} = 0.002$), \systemname scored similar to that of the baseline. P3 remarked that \participantquote{not a singular component by itself, but all components together do make me more engaged,}. P10 expressed a sense of active involvement in the task, stating that \participantquote{Yes. It may automatically detect my progress to make me more engaged in the task.}

\subsubsection{\systemname~ as a proactive AR assistant in everyday life.} 
Most participants agreed that \systemname~has the potential to be generalized to everyday scenarios (Q11: $p=0.277$ and $p_{non\_inferiority} = 0.005$). P9 said that \participantquote{maybe when we need to assemble furniture, instead of going through the manual back and forth all the time, we can just have this system to guide us.} 
Furthermore, most participants acknowledged that they would not need additional training to use the system (Q7: $p=0.179$ and $p_{non\_inferiority} = 0.001$), suggesting possible applications for more general purposes. With some training, as P10 mentioned, \participantquote{(The system can be used for) learning to complete a difficult task.}

\input{tables/sus}

\input{tables/tlx}

\begin{figure}[t]
\includegraphics[width=0.9\linewidth]{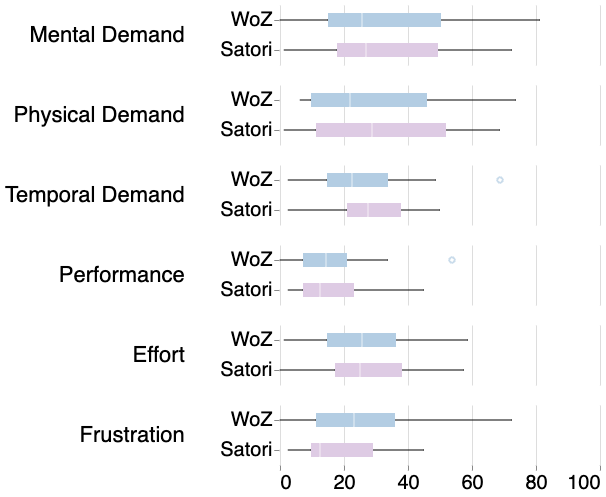}
\caption{The box plot of NASA-TLX results illustrates the distribution of cognitive load ratings across six dimensions: Mental Demand, Physical Demand, Temporal Demand, Performance, Effort, and Frustration. Each box represents the interquartile range (IQR) with the median marked by a horizontal line, showing the variability and central tendency of participants' workload ratings for both systems. The comparison highlights differences in perceived workload between the WoZ and Satori conditions, providing insights into the effectiveness and usability of each approach.}
\centering
\label{fig:tlx}
\end{figure}

\subsection{NASA TLX Result on cognitive load}
\revision{We found no significant difference between~\systemname~and~\baseline~on all TLX measures. Detailed analysis within the six sub-categories of NASA TLX revealed no significant difference among the six subcategories of NASA TLX between the two conditions, see Table~\ref{tab:tlx} and Figure~\ref{fig:tlx} for details.}

%% file: tables/desire.tex
\begin{table}[ht]
    \centering
    \begin{tabular}{ccc}
    \toprule
    Dataset & GTEA & Our Dataset \\ \midrule
    Satori & 100.00 & 100.00 \\
    \bottomrule
    \end{tabular}
    \caption{Desire inference includes understanding high-level task goals. We evaluated this module using the GTEA dataset and our dataset, which is annotated by three experimenters. Satori achieved a balanced accuracy of 100$\%$ on both datasets}
    \label{tab:desire-prediction}
\end{table}

%% file: tables/belief.tex






\begin{table}[ht]
\centering
\begin{tabular}{ccc}
\toprule

Dataset & EgoTaskQA & Our Dataset                    \\
Task    & Scene Understanding        & Object Understanding \\ \hline
HCRN    & 69.50      & N/A                            \\
Satori  & 66.50      & 57.90     \\
\bottomrule

\end{tabular}

\caption{
Belief inference includes scene understanding and task-relevant object understanding (object labels, locations) and their interaction history with the user. As for the evaluation, the goal is to understand the reasoning capability for scene understanding and object understanding. We evaluated this module using the EgoTaskQA dataset and our dataset to compare with the HCRN model. The EgoTaskQA dataset consists of questions about humans' understanding of the scene and the model's understanding of humans' beliefs. For our dataset, three experimenters annotated the highlighted object labels, locations, and interaction states separately. Satori reached a similar accuracy (66.50$\%$) to that of the HCRN model (69.50$\%$).}
\label{tab:belief-scene-prediction}

\end{table}

%% file: tables/intention.tex



\begin{table}[ht]
\centering
\begin{tabular}{lrrrrrr}
\toprule \multirow{2}{*}{L.A. Time}
 & \multicolumn{3}{c}{GTEA} & \multicolumn{3}{c}{Our Dataset} \\
\cmidrule(lr){2-4}\cmidrule(lr){5-7}
 & Recall & Prec. & F1 & Recall & Prec. & F1 \\
\midrule
0s      & 63.04 & 78.38 & 69.88 & \textbf{65.61} & 62.52 & 58.89 \\
1s & 54.35 & 75.76 & 63.29 & 55.00          & 48.40 & 46.06 \\
3s & 39.95 & 65.73 & 49.43 & 52.31          & 44.44 & 45.24 \\

\bottomrule
\end{tabular}
\caption{
This table shows the module-level evaluation of intention (action) forecast. L.A. Time refers to Look-Ahead Time, Prec. refers to Precision score. We evaluated our methods on the GTEA dataset and our dataset. Three experimenters annotated user action labels in our dataset. Aside from the settings Satori uses (Look-Ahead Time = 0s), we also present results for two other hypothetical conditions if we predict the action 1s or 3s earlier. For our settings, our methods reached 78.38$\%$ on the GTEA dataset and 62.52$\%$ on our dataset.}
\label{tab:intention-prediction}

\end{table}

%% file: tables/modality.tex

\begin{table}[ht]
\centering
\begin{tabular}{ccc}
\toprule
Task                       & Guidance Timing & Modality \\
\hline
Arranging Flowers          & 94.34             & 94.34     \\
Connecting NS & 79.49          & 74.15     \\
Room Cleaning              & 80.49             & 73.17     \\
Making Coffee              & 75.00             & 63.75     \\
Average                    & 81.69          & 75.12    \\
\bottomrule

\end{tabular}
\caption{
The table shows the modality prediction results using the user study videos for the four tasks: arranging flowers, connecting Nintendo Switch (NS), cleaning a room, and making coffee.
Three experimenters labeled the assistance appearances and compared them with the WoZ. Our methods reached an average of 75.12$\%$ when referring to the same assistance type as the designers'. The guidance timing columns show the holistic evaluation on whether Satori generated the proper assistance at the proper time without modality. }
\label{tab:modality-prediction}
\end{table}

%% file: tables/sus.tex
\begin{table*} [t]
\centering
\begin{tabular} {p{5cm} ccccccc} 
\hline
\multirow{2} {*} {Question}  & \multirow{2} {*} {Condition}   & \multirow{2} {*} {Mean}   & \multirow{2} {*} {95\% CI}   & \multicolumn{2} {c} {Vanilla}  &  \multicolumn{2} {c} {Non-Inferiority}   \\
& & & & W &$p$-value & W & $p$-value \\
\hline
\multirow{2} {5cm} {[Q1] I can easily comprehend content via text\/audio\/image guidance.}  & \systemname & 6.25 & [6.00, 6.75] & \cellcolor{white}  & \cellcolor{white}  & \cellcolor{lightgreen}  & \cellcolor{lightgreen}  \\
 & WoZ & 5.94 & [5.25, 6.50] & \multirow{-2} {*} {\cellcolor{white} 26.500}   & \multirow{-2} {*} {\cellcolor{white} 0.099}   & \multirow{-2} {*} {\cellcolor{lightgreen} 89.500}   &\multirow{-2} {*} {\cellcolor{lightgreen} 0.001}   \\
\hline
\multirow{2} {5cm} {[Q2] The guidance's content is helpful in completion the task.}  & \systemname & 6.22 & [5.75, 6.75] & \cellcolor{white}  & \cellcolor{white}  & \cellcolor{lightgreen}  & \cellcolor{lightgreen}  \\
 & WoZ & 5.80 & [5.50, 6.50] & \multirow{-2} {*} {\cellcolor{white} 26.000}   & \multirow{-2} {*} {\cellcolor{white} 0.094}   & \multirow{-2} {*} {\cellcolor{lightgreen} 131.000}   &\multirow{-2} {*} {\cellcolor{lightgreen} 0.000}   \\
\hline
\multirow{2} {5cm} {[Q3] I think the guidance appear at the right moment.}  & \systemname & 5.97 & [6.00, 6.50] & \cellcolor{white}  & \cellcolor{white}  & \cellcolor{lightgreen}  & \cellcolor{lightgreen}  \\
 & WoZ & 5.53 & [5.00, 6.25] & \multirow{-2} {*} {\cellcolor{white} 17.500}   & \multirow{-2} {*} {\cellcolor{white} 0.090}   & \multirow{-2} {*} {\cellcolor{lightgreen} 125.000}   &\multirow{-2} {*} {\cellcolor{lightgreen} 0.001}   \\
\hline
\multirow{2} {5cm} {[Q4] I found that the guidance accurately reflects my task intentions.}  & \systemname & 6.48 & [6.00, 7.00] & \cellcolor{lightgreen}  & \cellcolor{lightgreen}  & \cellcolor{lightgreen}  & \cellcolor{lightgreen}  \\
 & WoZ & 5.95 & [5.62, 6.50] & \multirow{-2} {*} {\cellcolor{lightgreen} 11.500}   & \multirow{-2} {*} {\cellcolor{lightgreen} 0.016}   & \multirow{-2} {*} {\cellcolor{lightgreen} 134.500}   &\multirow{-2} {*} {\cellcolor{lightgreen} 0.000}   \\
\hline
\multirow{2} {5cm} {[Q5] The guidance appears at a adequate location.}  & \systemname & 6.23 & [5.88, 6.75] & \cellcolor{lightgreen}  & \cellcolor{lightgreen}  & \cellcolor{lightgreen}  & \cellcolor{lightgreen}  \\
 & WoZ & 5.66 & [5.25, 6.25] & \multirow{-2} {*} {\cellcolor{lightgreen} 15.000}   & \multirow{-2} {*} {\cellcolor{lightgreen} 0.032}   & \multirow{-2} {*} {\cellcolor{lightgreen} 131.500}   &\multirow{-2} {*} {\cellcolor{lightgreen} 0.000}   \\
\hline
\multirow{2} {5cm} {[Q6] I am able to complete my work quickly using this system.}  & \systemname & 6.08 & [5.50, 6.62] & \cellcolor{white}  & \cellcolor{white}  & \cellcolor{lightgreen}  & \cellcolor{lightgreen}  \\
 & WoZ & 5.75 & [5.25, 6.50] & \multirow{-2} {*} {\cellcolor{white} 30.000}   & \multirow{-2} {*} {\cellcolor{white} 0.273}   & \multirow{-2} {*} {\cellcolor{lightgreen} 108.500}   &\multirow{-2} {*} {\cellcolor{lightgreen} 0.003}   \\
\hline
\multirow{2} {5cm} {[Q7] It was easy to learn to use this system.}  & \systemname & 6.48 & [6.00, 7.00] & \cellcolor{white}  & \cellcolor{white}  & \cellcolor{lightgreen}  & \cellcolor{lightgreen}  \\
 & WoZ & 6.06 & [5.75, 7.00] & \multirow{-2} {*} {\cellcolor{white} 22.000}   & \multirow{-2} {*} {\cellcolor{white} 0.179}   & \multirow{-2} {*} {\cellcolor{lightgreen} 103.500}   &\multirow{-2} {*} {\cellcolor{lightgreen} 0.001}   \\
\hline
\multirow{2} {5cm} {[Q8] How engaged I am using the system?}  & \systemname & 6.16 & [5.88, 6.50] & \cellcolor{white}  & \cellcolor{white}  & \cellcolor{lightgreen}  & \cellcolor{lightgreen}  \\
 & WoZ & 5.75 & [5.38, 6.50] & \multirow{-2} {*} {\cellcolor{white} 20.500}   & \multirow{-2} {*} {\cellcolor{white} 0.145}   & \multirow{-2} {*} {\cellcolor{lightgreen} 109.500}   &\multirow{-2} {*} {\cellcolor{lightgreen} 0.002}   \\
\hline
\multirow{2} {5cm} {[Q9] The system's guidance matches the context.}  & \systemname & 6.27 & [6.00, 6.75] & \cellcolor{white}  & \cellcolor{white}  & \cellcolor{lightgreen}  & \cellcolor{lightgreen}  \\
 & WoZ & 6.05 & [5.62, 7.00] & \multirow{-2} {*} {\cellcolor{white} 32.500}   & \multirow{-2} {*} {\cellcolor{white} 0.357}   & \multirow{-2} {*} {\cellcolor{lightgreen} 91.000}   &\multirow{-2} {*} {\cellcolor{lightgreen} 0.001}   \\
\hline
\multirow{2} {5cm} {[Q10] Overall, the system's guidance frequency is suitable.}  & \systemname & 6.30 & [5.75, 6.75] & \cellcolor{white}  & \cellcolor{white}  & \cellcolor{lightgreen}  & \cellcolor{lightgreen}  \\
 & WoZ & 5.92 & [5.75, 6.50] & \multirow{-2} {*} {\cellcolor{white} 30.000}   & \multirow{-2} {*} {\cellcolor{white} 0.156}   & \multirow{-2} {*} {\cellcolor{lightgreen} 97.500}   &\multirow{-2} {*} {\cellcolor{lightgreen} 0.002}   \\
\hline
\multirow{2} {5cm} {[Q11] Overall, I think the system can be beneficial in my everyday life.}  & \systemname & 5.94 & [5.50, 6.50] & \cellcolor{white}  & \cellcolor{white}  & \cellcolor{lightgreen}  & \cellcolor{lightgreen}  \\
 & WoZ & 5.58 & [5.25, 6.25] & \multirow{-2} {*} {\cellcolor{white} 30.000}   & \multirow{-2} {*} {\cellcolor{white} 0.277}   & \multirow{-2} {*} {\cellcolor{lightgreen} 105.500}   &\multirow{-2} {*} {\cellcolor{lightgreen} 0.005}   \\
\hline
\end{tabular} 
\vspace{7 pt} 
\caption{The table summarizes the mean scores and 95\% confidence intervals (CI) for each system (our Satori system and WoZ designed by the AR designer) across usability scale questions using non-inferiority tests. The ``Vanilla'' columns provide the Wilcoxon signed-rank test results (W statistic and p-values) for significant differences between systems. The ``Non-Inferiority'' columns show W statistics and p-values testing if Satori's performance is non-inferior to WoZ within a set margin. The highlighted cells indicate established non-inferiority, suggesting that Satori performs comparably or better than WoZ over system performance and usability.} 
\label{tab:sus} 
\end{table*} 

%% file: tables/tlx.tex
\begin{table*}[ht]
\centering
\begin{tabular}{p{5cm}ccccccc}
\hline
\multirow{2}{*}{Question} & \multirow{2}{*}{Condition}  & \multirow{2}{*}{Mean}  & \multirow{2}{*}{95\% CI}  & \multicolumn{2}{c}{Vanilla} &  \multicolumn{2}{c}{Non-Inferiority}  \\
& & & & W &$p$-value & W & $p$-value \\
\hline
\multirow{2}{5cm}{Mental Demand} & Satori & 34.06 & [17.50, 43.81] & \cellcolor{white} & \cellcolor{white} & \cellcolor{white} & \cellcolor{white} \\
 & WoZ & 33.12 & [16.25, 50.00] & \multirow{-2}{*}{\cellcolor{white}60.500}  & \multirow{-2}{*}{\cellcolor{white}0.744}  & \multirow{-2}{*}{\cellcolor{white}78.000}  &\multirow{-2}{*}{\cellcolor{white}0.316}  \\
\hline
\multirow{2}{5cm}{Physical Demand} & Satori & 32.34 & [11.25, 50.00] & \cellcolor{white} & \cellcolor{white} & \cellcolor{white} & \cellcolor{white} \\
 & WoZ & 30.47 & [10.00, 43.75] & \multirow{-2}{*}{\cellcolor{white}55.000}  & \multirow{-2}{*}{\cellcolor{white}0.776}  & \multirow{-2}{*}{\cellcolor{white}80.000}  &\multirow{-2}{*}{\cellcolor{white}0.281}  \\
\hline
\multirow{2}{5cm}{Temporal Demand} & Satori & 28.52 & [21.25, 37.50] & \cellcolor{white} & \cellcolor{white} & \cellcolor{white} & \cellcolor{white} \\
 & WoZ & 26.41 & [15.00, 32.50] & \multirow{-2}{*}{\cellcolor{white}46.000}  & \multirow{-2}{*}{\cellcolor{white}0.274}  & \multirow{-2}{*}{\cellcolor{white}65.000}  &\multirow{-2}{*}{\cellcolor{white}0.388}  \\
\hline
\multirow{2}{5cm}{Performance} & Satori & 16.17 & [7.50, 21.25] & \cellcolor{white} & \cellcolor{white} & \cellcolor{lightgreen} & \cellcolor{lightgreen} \\
 & WoZ & 17.27 & [7.50, 20.00] & \multirow{-2}{*}{\cellcolor{white}44.000}  & \multirow{-2}{*}{\cellcolor{white}0.593}  & \multirow{-2}{*}{\cellcolor{lightgreen}95.500}  &\multirow{-2}{*}{\cellcolor{lightgreen}0.022}  \\
\hline
\multirow{2}{5cm}{Effort} & Satori & 28.20 & [17.50, 37.50] & \cellcolor{white} & \cellcolor{white} & \cellcolor{white} & \cellcolor{white} \\
 & WoZ & 26.02 & [15.00, 36.25] & \multirow{-2}{*}{\cellcolor{white}52.500}  & \multirow{-2}{*}{\cellcolor{white}0.464}  & \multirow{-2}{*}{\cellcolor{white}58.500}  &\multirow{-2}{*}{\cellcolor{white}0.353}  \\
\hline
\multirow{2}{5cm}{Frustration} & Satori & 19.84 & [10.00, 28.75] & \cellcolor{white} & \cellcolor{white} & \cellcolor{lightgreen} & \cellcolor{lightgreen} \\
 & WoZ & 26.95 & [11.25, 34.38] & \multirow{-2}{*}{\cellcolor{white}41.500}  & \multirow{-2}{*}{\cellcolor{white}0.175}  & \multirow{-2}{*}{\cellcolor{lightgreen}126.000}  &\multirow{-2}{*}{\cellcolor{lightgreen}0.001}  \\
\hline
\end{tabular}
\caption{This table shows the results for NASA TLX questions and non-inferiority tests using the mean scores and 95\% confidence intervals (CI) for Satori and WoZ systems across six dimensions: Mental Demand, Physical Demand, Temporal Demand, Performance, Effort, and Frustration. The Vanilla Wilcoxon signed-rank test results and non-inferiority tests (highlighted in green) indicate whether the Satori system performs comparably or better than the WoZ system in terms of cognitive load.}
\label{tab:tlx}
\end{table*}

%% file: sections/06_discussion.tex
\section{Discussion}
\subsection{Toward Proactive AR Assistance}
Our \systemname~system represents an early attempt to provide appropriate assistance at the right time. The findings in timing, comprehensibility, and effectiveness all demonstrated that \systemname~performs similarly to AR assistance created by AR designers, marking a successful \revision{proof-of-concept of integrating the BDI model into} AR assistance \revision{(RQ1 and RQ2)}. Many participants reported that they could not tell which condition they were using. This is due to the joint implementation of \textit{belief} (task prediction, step-by-step instructions), \textit{desire} (action prediction), and intention (content and timing prediction). During the study, we observed that \systemname's implementation prevents duplicated task or step prediction, which was a main challenge in the initial testing as it confuses the users. None of the participants reported feeling confused by repetition in the dynamically updated AR assistance. 

\revision{Collaborating with a proactive assistant to complete a task could be a form of human-AI collaboration. Participants were generally positive about the model confirming their intentions, as P5 put it: \participantquote{It (\systemname) gives me the impression that the machine understands what I'm doing, making its instructions feel trustworthy}. However, when the system failed or did not predict the correct step, we noticed that users did not always tell the model what to do by selecting the ``No'' or ``Cancel'' button to prompt the model to retry; instead, they sometimes proceeded off the script, improvising next steps until the system picked up on their actions. Notably, none of the users completely ``abandoned'' collaboration with the AI, as they still periodically checked to see if the assistant had caught up (indicating some degree of trust). This differs from our initial user testing when the AR assistance often failed to track actions or tasks, resulting in users completely relying on their own discretion and ignoring what the system said (indicating low trust). This suggests that human-AI collaboration might need systems to pass a minimum usability or performance threshold for users to really benefit from their assistance.}

\revision{\subsection{\systemname~and baseline in guiding users}
Given that the \baseline~was designed by AR designers who carefully considered the timing and content of the assistance for the four scenarios, \systemname~ achieved promising outcomes in comprehensibility, efficiency, and cognitive load without extensive manual effort. This suggests that AR assistance and guidance can be partially automated and benefit from reusable components (RQ3). } 

\revision{Participants agreed that both \systemname~ and the baseline systems provide clear and efficient instructions (See \revision{Table}~\ref{tab:sus}), as P8 noted that \participantquote{...the instructions were very clear in both the things...} and P13 said that \participantquote{It (Satori) is effective at helping me at all tasks, and the UI is clear for both}. However, the baseline seems to perform faster (P4, P2), with animations that are ``on-point'' (P6). However, not having animations in \systemname~ did not cause tremendous issues in understanding the task. This may be because the arrows \systemname~ dynamically uses to point toward actions are considered ``effective'' by participants (P2, P1, P6).} 

\revision{\subsection{\systemname's~ prediction performance}
Satori's timing evaluation (i.e., intention prediction) scored somewhat highly in accuracy on both the GTEA data and our datasets. Although its predictive capacity is imperfect, subjective ratings suggest that users do not notice the difference in practice. This aligns with our in-study observations, where we assessed that the actual effect of the prediction errors is that the AR assistance will appear a few seconds earlier or later at the beginning and end of each step. The confirmation page prevented the timing errors at the beginning of each step by aligning users' intentions and the task at hand. As for errors at the end of each step, they occasionally caused participants to focus on the assistance a few seconds after their actions were completed. As a result, the range of timing errors was well-tolerated and did not impact the actual performance.}

\revision{Despite Satori's limited predictive accuracy, it did not ultimately impact user experience. The modality's fairly high recall (75.12$\%$) is compared with the AR designer's design. Even if the modality does not compare, the effect did not seem to be one of misleading the user or undermining their experience. Since text instructions are always displayed, as recommended by the groups in the formative study, Satori's modality prediction alternates among images, image sequences, audio, and tools. For example, when the AR designer used audio but Satori presented an image, the actual effect of the error was challenging to measure per step, but most participants agreed that Satori's multimodal content presentation was easily understandable (Q1). As a result, holistic system evaluation from the user's perspective were important since ensuring practical user experience is an essential part of AR assistance.}

\subsection{Advanced methods of user modelling in AR}
\revision{Our implementation of the BDI model provides a rule-based framework for determining assistance timing, modality, and content, relying on deterministic predictions of actions, goals, and tasks. Advanced modeling methods, such as the COBO framework \cite{yu2022optimizing}, introduce more sophisticated techniques for BDI representation; they can, for example, explicitly account for both benefits and costs to optimize assistance timing. Benefits might include reduced overall task completion time, while costs could involve cognitive interruptions, processing delays, and disruptions to task flow. } 

\revision{A more quantitative way to model BDI could entail a continuous confidence score to capture the probabilistic nature of actions, tasks, and goals. For instance, our early testing suggests that suggestions that align with the user's intention are beneficial, while repeated suggestions are generally discouraging; yet, in challenging tasks, users may appreciate repeated guidance for steps they have already completed.  To gain additional precision, one strategy might be to decode neural signals to provide a direct measure of intention. For example, \cite{nguyen2023modeling} proposes multimodal methods that integrate EEG and EMG signals to decode mental states. However, such an approach may encounter further challenges with accuracy and portability for everyday use.  }


\subsection{Challenges, improvements, and generalizability}
Despite our efforts, there is still substantial room for system improvement, as well as for work toward more general proactive AR assistance. The current system still has a latency of about 2-3 seconds (limiting application to non-rapid performance), works best on pre-defined tasks and can be affected by the limited FoV of the AR device. On the other hand, however, \systemname provides an example of a system where modalities and use scenarios can be scaled. We describe insights below:

\subsubsection{Interaction latency}
Even though~\systemname~manages to reduce user wait time via the early forecasting mechanism, the latency was still sometimes detectable to participants. P8 mentioned, \participantquote{Most of the time the system knows what I have done in the past step eventually, but I wish it could be more responsive so I don't need to wait for the system to recognize what I have done.} The action finish detection module is the main source of this latency, as it shares GPU resources with the goal prediction module; the data transfer between the GPU and CPU can sometimes lead to latency in the action finish detection module. We could eliminate this latency by adding more computing resources to avoid memory sharing between different modules.

\subsubsection{From pre-defined to adaptive tasks}
\revision{For the task prediction used in \systemname, we began with a task planner that matches the tasks to be executed with entries from an existing database, which is built from manual input. Although this is not the main focus of this paper, enabling automatic task guidance authoring could improve the system's scalability. Automatic content input poses additional challenges in the \textit{how to} efficiently search, dissect, and format different parts (e.g., image, text, animation) of the instructions into multimodal AR assistance, which is beyond this paper's scope. That said, \systemname's implementation does not rely on a specific label set or manual configuration, making it adaptable to most new tasks regardless of whether they are pre-made or not.  }

\subsubsection{Constraints in FoV}
\revision{Despite improvement to horizontal FoV, HoloLen2's vertical FoV is only around 30 degrees, meaning it sometimes clips off important contextual information about users' hand actions. Video streams from the HoloLens's official Mixed Reality Capture showed that users' hands are not always in view and user interactions are sometimes not visible to the AR device}. The issue is further exacerbated when the user moves or rotates their head, or looks up. Guidance mismatch and prediction error often follow. Although \systemname's user confirmation step allows users to manually adjust or correct the system's understanding, frequent confirmation could lead to decreased interaction flow and a sub-optimal experience. Despite model improvement being a crucial factor, using a device with a larger vertical FoV might make an instant improvement to the overall experience. 

Another possible improvement would entail additional environmental sensing capabilities, such as better cameras or integrating third-person view. Additional sensors could provide a more holistic understanding of the environment. This would allow the AR assistant to offer more precise and contextually appropriate guidance and \revision{reduce errors caused by missing information}. A third-person view could provide contextual information when the user's first-person view is obstructed or occluded~\cite{lee2019optical}.

\subsubsection{\revision{A broader range of assistance modalities could cater to diverse AR assistance needs.}} The spectrum of human task activities is vast, ranging from highly cognitive tasks to more physical ones~\cite{costley2017video}. \revision{\systemname's current output modalities provide feedback in basic text, image, and audio, but do not harness more advanced feedback modes like visualizations, assistive tools, or animation. Expanding assistance modalities could present new opportunities for scalability and adaptability}, such as auditory warnings, dynamic animations, or AI-supported navigating or counting tools. These modalities could further bridge the gap between virtual guidance and real-world task execution by more closely aligning assistance with specific task demands.

\subsubsection{Future development for collaborative interactions and social dynamics.}
Humans often perform tasks in group settings, where individual intentions are influenced by social interactions, such as communication, collaboration, negotiation, and working toward shared goals. During our interview, participants suggested that the future proactive AR assistant should be capable of recognizing social interactions and multi-user scenarios, adapting the assistance's content, timing, and frequency to support the group. This view reflects similar results from non-AR studies that have highlighted the importance of collaborative interaction for AI assistance~\cite{DBLP:conf/acl/WuZS024}.

\subsection{Limitation and Future Directions}
The primary limitation lies in incomplete prediction of user goals (i.e., surrounding objects, history, and actions). In psychology, human beliefs are highly complex and nuanced, and our current implementation only partially captures this complexity. Research in cognitive psychology suggests that human beliefs are influenced by personal experiences, social influences, and cognitive biases~\cite{kahneman2013prospect, ross2011person}. Thus, our system may benefit from incorporating more sophisticated models and diving into more methods from interdisciplinary research on neuroscience and decision-making.

Additionally, our implementation primarily relies on the GPT model, which suffers from network latency, content load, and privacy issues. Although we experimented with LLaVA in our early testings, the results were not satisfactory. Future work may consider balancing the computation between the online LLM models and localized models to mitigate the above issues. 

\revision{Currently our work only supports the domain of everyday tasks. Our implementation using LLM has limited efficacy for domain-specific applications. This is critical since AR assistance is widely used in specific domains like industry, medicine, and education, and its goal is to improve efficiency and reduce human errors. Future work should examine proactive AR assistance in these domains, as a successful application could support life-saving work and safety. }

%% file: sections/07_conclusion.tex
\section{Conclusion}
We presented \systemname, a proactive AR assistant system that integrates the concepts from the belief-desire-intention model with fusion architecture comprised of LLMs and local vision models to achieve timely, context-specific, multimodal AR assistance. Our research expands the bounds of the field by presenting a proactive assistance that uses user actions, task goals, environmental context, and scene objects to automatically provide step-by-step AR assistance. Two formative studies involving 12 experts identified four design requirements for creating proactive AR assistance, emphasizing the importance of understanding human actions, surrounding objects, and task context. Integrating with concepts from the BDI model, \systemname~is capable of automatically providing step-by-step instructions that respond to users' task progress. An empirical study with 16 participants demonstrated that \systemname~performs comparably to designer-created AR assistance in task guidance for timeliness, content comprehensibility, usefulness, and efficacy.  The results indicate that by capturing both user intentions and semantic context, \systemname~ could be used to reduce repetitive creation on similar AR assistance and increase generalizability and reusability, potentially improving the scalability issues faced by existing AR assistance. Our work opens new human-AI collaborative AR experiences for a range of tasks.